\documentclass[preprint,aps]{revtex4}
\usepackage{mathrsfs}
\usepackage{graphicx}
\usepackage{array,threeparttable}

\newcommand{\PreserveBackslash}[1]{\let\temp=\\#1\let\\=\temp}
\newcolumntype{C}[1]{>{\PreserveBackslash\centering}p{#1}}
\newcolumntype{R}[1]{>{\PreserveBackslash\raggedleft}p{#1}}
\newcolumntype{L}[1]{>{\PreserveBackslash\raggedright}p{#1}}

\begin{document}

\title{Electronic Structure and Superconductivity of FeSe-Related Superconductors}

\author{Xu Liu$^1$, Lin Zhao$^1$, Shaolong He$^1$, Junfeng He$^1$, Defa Liu$^1$, Daixiang Mou$^1$, Bing Shen$^1$, Yong Hu$^1$, Jianwei Huang$^1$ and X. J. Zhou$^{1,2}$$^,$$^*$}

\affiliation{
\\$^{1}$National Lab for Superconductivity, Beijing National Laboratory for Condensed Matter Physics, Institute of Physics,
Chinese Academy of Sciences, Beijing 100190, China
\\$^{2}$Collaborative Innovation Center of Quantum Matter, Beijing, China
}
\date{November 24, 2014}

\begin{abstract}

The FeSe superconductor and its related systems have attracted much attention in the iron-based superconductors owing to their simple crystal structure and peculiar electronic and physical properties. The bulk FeSe superconductor has a superconducting transition temperature (T$_c$) of $\sim$8 K; it can be dramatically enhanced to 37 K at high pressure. On the other hand, its cousin system, FeTe, possesses a unique antiferromagnetic ground state but is non-superconducting. Substitution of Se by Te in the FeSe superconductor results in an enhancement of T$_c$ up to 14.5 K and superconductivity can persist over a large composition range in the Fe(Se,Te) system. Intercalation of the FeSe superconductor leads to the discovery of the A$_x$Fe$_{2-y}$Se$_2$ (A=K, Cs and Tl) system that exhibits a T$_c$ higher than 30 K and a unique electronic structure of the superconducting phase. The latest report of possible high temperature superconductivity in the single-layer FeSe/SrTiO$_3$ films with a T$_c$ above 65 K has generated much excitement in the community. This pioneering work opens a door for interface superconductivity to explore for high T$_c$ superconductors. The distinct electronic structure and superconducting gap, layer-dependent behavior and insulator-superconductor transition of the FeSe/SrTiO$_3$ films provide critical information in understanding the superconductivity mechanism of the iron-based superconductors. In this paper, we present a brief review on the investigation of the electronic structure and superconductivity of the FeSe superconductor and related systems, with a particular focus on the FeSe films.

\end{abstract}


\maketitle

\tableofcontents
\newpage

\section{Introduction}

The iron-based superconductors discovered in 2008\cite{Kamihara} represent the second class of high temperature superconductors after the discovery of the first class of high-T$_c$ cuprate superconductors in 1986\cite{Bednorz}. The superconducting transition temperature (T$_c$) has reached $\sim$55 K\cite{XHChen,NLWang,ZARenSm,ZAXu} that is beyond the generally-believed McMillan limit of the conventional superconductors. Indications of even higher T$_c$ have emerged in the single-layer FeSe films\cite{QYWangCPL,DFLiu,SLHe,STan}.   Since the discovery of the cuprate superconductors, understanding the high temperature superconductivity mechanism remains a prominent and challenging task facing the condensed matter physics community.  The discovery of the iron-based superconductors provides an opportunity to compare and contrast with the cuprate superconductors that may lead to uncover the clue of high temperature superconductivity. Great progress has been made in materials preparation, experimental investigation, and theoretical understanding of the iron-based superconductors\cite{KIshida,Johnston,Paglinone,Stewart,FWang,Hirschfeld,DXMouReview,PCDai,Chubukov,Dagotto}.

So far, several families of the iron-based superconductors have been discovered that can be mainly categorized into `11'\cite{MKWu}, `111'\cite{XCWang}, `122'\cite{MRotter} and `1111'\cite{Kamihara,XHChen,NLWang,ZARenSm} systems according to their crystal structure (Fig. 1)\cite{Paglinone}. Similar to the cuprate superconductors, the iron-based superconductors are quasi-two-dimensional in their crystal structure. The FePn (Pn=As or Se) layer is an essential building block that is believed to be responsible for the superconductivity in the iron-based superconductors. Different from the cuprate superconductors where the CuO$_2$ plane is basically co-planar, the FePn (Pn=As or Se) unit consists of three layers with the central Fe layer sandwiched in between two adjacent Pn (Pn=As or Se) layers. This results in the doubling of the unit cell in the iron-based superconductors and the folding of the corresponding electronic structure (Fig. 2). Most significantly, different from the cuprate superconductors where the low-energy physics is mainly dominated by the single Cu d$_{x^2-y^2}$ orbital, in the iron-based superconductors, all the five Fe 3d orbitals participate in the low-energy electronic structures\cite{FJMaFPC,TYildirim,FJMaPRB,FJMaPRL}.  Generally, there are multiple bands crossing the Fermi level that form hole-like Fermi surface sheets near the Brillouin zone center and electron-like Fermi surface sheets near the zone corner\cite{FChen,YLubashevsky,JMaletz,AAKordyuk,LZhao,GDLiu,HYLiuPRB,HYLiuPRL}. The multiple-orbital nature (Fig. 2)\cite{SGraser,YZhangorbital} plays an important role in understanding the physical properties and superconductivity mechanism in the iron-based superconductors.


Among all the iron-based superconductors, the FeSe superconductor and its related systems have gained particular attention due to their simple crystal structure and peculiar electronic and physical properties. The FeSe superconductor has a simple crystal structure consisting of the FeSe layer that is the essential building block in the iron-based superconductors; such a simple structure is ideal for theoretical and experimental study of the superconductivity mechanism. The bulk FeSe superconductor has a superconducting transition temperature (T$_c$) of $\sim$8 K\cite{MKWu}; it can be dramatically enhanced to 37 K at high pressure\cite{SMedvedev}. On the other hand, its cousin system, FeTe, possesses a unique antiferromagnetic ground state but is non-superconducting\cite{WBao}. Substitution of Se by Te in the FeSe superconductor results in an enhancement of T$_c$ up to 14.5 K and superconductivity can persist over a large composition range in the Fe(Se,Te) system\cite{MHFangPRB,Nkatayama}. Intercalation of the FeSe superconductor leads to the discovery of the A$_x$Fe$_{2-y}$Se$_2$ (A=K, Cs and Tl) system that exhibits a T$_c$ higher than 30 K\cite{JGuo,MFang,MHFangEPL,XLChenSR} and unique electronic structure of the superconducting phase\cite{DLFengIFS,HDingIFS,DXMouReview}. The latest report of possible high temperature superconductivity in the single-layer FeSe/SrTiO$_3$ films with a T$_c$ above 65 K has generated much excitement in the community\cite{QYWangCPL,DFLiu,SLHe,STan}. This pioneering work opens a door for interface superconductivity to explore for high T$_c$ superconductors\cite{QYWangCPL}. The distinct electronic structure and superconducting gap, layer-dependent behavior and insulator-superconductor transition of the FeSe/SrTiO$_3$ films provide critical information in understanding the superconductivity mechanism of the iron-based superconductors.

In this paper, we will present a brief review on the investigation of the electronic structure and superconductivity of the FeSe superconductor and related systems. We will put particular focus on the FeSe films that is an exciting and fast-growing field. The paper is organized as follows: In Section 2, bulk FeSe and related materials are first introduced. In Section 3, we will discuss the electronic properties of the FeSe films, including Fermi surface, band structure, gap symmetry and the evolution of electron structure with annealing. The implications and theoretical understandings are presented in Section 4.  In Section 5, we end with further issues to be investigated and  a future perspective.


\section{Electronic Structure of Bulk F\MakeLowercase{e}(S\MakeLowercase{e},T\MakeLowercase{e}) System and Intercalated F\MakeLowercase{e}S\MakeLowercase{e}}

\subsection{Bulk Fe(Se,Te)}






Bulk FeSe superconductor with a  T$_c$ of 8 K was first discovered in 2008 in a tetragonal phase $\beta$-FeSe with PbO-structure at ambient pressure (Fig. 3a)\cite{MKWu}.  It has been found that excess Fe is inevitable to stabilize the crystal structure of Fe$_{1+\delta}$Se and the superconductivity is very sensitive to its stoichiometry ($\delta$)\cite{TMMcQueen}. The FeSe superconductor exhibits a dramatic pressure dependence (Fig.3c), in particular, its T$_c$ can be enhanced to 36.7 K under high pressure\cite{SMedvedev}. At ambient pressure, FeSe superconductor undergoes a structural transition from tetragonal to orthorhombic around 90 K but without a magnetic transition\cite{YMizuguchi,TMMcQueen}. This is distinct from many other iron-based parent compounds where the structural transition is usually accompanied by a magnetic phase transition\cite{PCDaiTransition,JZhaoTransition,YChen,JZhao1111,JZhao122,AIGoldman,QHuang}. However, under high pressure, static magnetic order is observed in superconducting FeSe\cite{MBendele}. Short range spin fluctuation has also been reported in FeSe superconductor that is enhanced under high pressure\cite{Imai}.

As Se is gradually replaced by Te in FeSe, superconductivity with a maximum T$_c$ around 14.5 K can be observed in a large composition range (x) in the Fe$_{1+\delta}$Se$_{1-x}$Te$_x$ system (Fig. 3d)\cite{MHFangPRB,NTakayama}. Again, the excess Fe content ($\delta$) has a big effect on the physical properties and superconductivity in this case\cite{TMMcQueen}. The excess Fe is found to be located at the interstitial sites\cite{WBao}.   Fe$_{1+\delta}$Te is present at the other end of the Fe$_{1+\delta}$(Se$_{1-x}$Te$_x$) phase diagram (Fig. 3d). It undergoes a phase transition between 60-75 K\cite{MHFangPRB,GFChenFeTe} and no superconductivity is observed in the bulk Fe$_{1+\delta}$Te. Signature of superconductivity is reported in the FeTe films but is likely due to the incorporation of oxygens\cite{LXCao,YFNie,MZheng}.  At low temperature, Fe$_{1+\delta}$Te shows an antiferromagnetic order of bi-collinear structure (Fig. 3b, upper panel) that is distinct from the usual collinear magnetic structure observed in many other parent compounds like BaFe$_2$As$_2$ (Fig. 3b, lower panel)\cite{WBao,SLLiFeTe}.

The band structure and Fermi surface of the bulk FeSe superconductor (Fig. 4a and 4b) from the band structure calculations show similar behaviors as other iron-based superconductors, i.e., the low energy electronic states originate mainly from the iron 3d orbitals, and there are two hole-like Fermi surface sheets at the zone center and two intersecting electron-like Fermi surface sheets around the zone corner\cite{ASubedi}. Direct ARPES measurements on the FeSe superconductors has been hindered due to the difficulty in obtaining FeSe single crystals. With the latest progress on growing high-quality FeSe single crystals\cite{JYLin,AEBoehmer,KKhuynh,FZhou}, several ARPES measurements on the FeSe superconductor have become available\cite{JMaletz,KNakayama,TShimojima}.


The ARPES results on FeSe single crystals reported so far give a basically consistent picture\cite{JMaletz,KNakayama,TShimojima}, as exemplified in Fig. 4.  First, at low temperature below the structural transition ($\sim$90 K),  the $\Gamma$ point is dominated by two hole-like bands (Fig. 4f), while near the M point, an electron-like band crossing the Fermi level and a hole-like band at higher binding energy are observed. Second, the measured electronic structures show a significant difference from the band structure calculations; the bands undergo a pronounced orbital-dependent renormalization. Third,  there is an obvious change of the electronic structure across the structural phase transition\cite{KNakayama,TShimojima}. An energy splitting of bands up to 50 meV around the M point is observed below the structural transition.  This is interpreted as the splitting of the  d$_{yz}$ and d$_{xz}$  orbital bands (Fig. 4e), indicating the orbital ordering. In particular, the Fermi energy for both the hole-like bands near the zone center and the electron-like band near M is rather small. These are consistent with the quantum oscillation measurements\cite{TTerashima,AAudouard} and STM measurement\cite{SKasahara} of the FeSe crystals. The comparable energy scale of the Fermi energy, superconducting gap, and Zeeman energy in FeSe superconductor also provides a window to look into the BCS-BEC crossover regime\cite{SKasahara}.


FeTe\cite{MHFangPRB,GFChenFeTe},  as a cousin of FeSe, shows quite different behaviors. It is non-superconducting and undergoes a strutural/magnetic transition around T$_{MS}$=70 K\cite{MHFangPRB,GFChenFeTe} (Fig. 5e). It is semiconductor-like above the transition while becomes metallic below the transition (Fig. 5e).  Initial ARPES measurements reveal a pair of nearly electron-hole compensated Fermi pockets, strong Fermi velocity renormalization, and an absence of a spin-density-wave gap\cite{YXia}. The measured electronic structure is qualitatively similar to other iron-pnictides although the antiferromagnetic structure in FeTe is different and 45 degrees rotated when compared with that in other iron-pnictide parent compounds. This is not consistent with the Fermi surface nesting picture for the formation of spin-density-wave (SDW) below the transition temperature.  Further work studied the Fermi surface and band structure of FeTe both in the paramagnetic state and the SDW state\cite{YZhangFeTe}. As shown in Fig. 5, above the magnetic transition temperature, the spectral weight mainly concentrates near $\Gamma$ point and M point (Fig.5a) (as noted by \cite{ZKLiuFeTe}, there may be a 45 degree rotation on the Brillouin zone definition in \cite{YZhangFeTe}). The $\Gamma$ point is dominated by two hole-like bands while there are two electron-like bands near the M point.  Upon decreasing the temperature below T$_{MS}$, the spectral weight distribution experiences an obvious change. The spectral weight around M and X point is suppressed  that is interpreted as the spectral weight transfer from low binding energy to high binding energy\cite{YZhangFeTe}.  Detailed temperature-dependent ARPES measurements on FeTe reveal coherence quasipartical peak near the Fermi level below the transition, accompanied with a hump structure located at higher binding energy and a dip in between these two features (Fig. 5h and i)\cite{YZhangFeTe,ZKLiuFeTe}. Such a ``peak-dip-hump" structure bears a strong resemblance to that observed in manganites and the heavily underdoped cuprates that can be explained in terms of polaron formation\cite{ZKLiuFeTe}. The hump feature is interpreted as from incoherent excitation of electron strongly coupled to bosons while the qusipartical feature is related with coherent polaron motion. In this picture, the coherent polaron motion at low temperature may  explain the metallic transport in the spin-density-wave state\cite{ZKLiuFeTe}.


Extensive ARPES measurements have been carried out on Fe(Se,Te) system to investigate its electronic structure and superconducting gap\cite{KNakayamaFST,FChen,ATamai,YLubashevsky,HMiaoFeTeSe,KOkazakiFeTeSe,KOkazakiSRep,EIeki}. Typical results are summarized in Fig. 6.  There are three hole-like bands around the $\Gamma$ point with two bands crossing the Fermi level (labeled as $\beta$ and $\gamma$) and the third band (labeled as $\alpha$) barely touching the Fermi level, while there are two electron-like bands around M point in Fe(Se,Te) (Fig. 6c-e)\cite{FChen}. The measured Fermi surface is sketched in Fig.6f.  Taking advantages of the photoemission matrix element effect by performing ARPES measurements under different polarization geometry, the $\alpha$ band is assigned as from d$_{xz}$/d$_{yz}$ orbitals, the $\beta$ band is mainly from d$_{xz}, d_{yz}$ and some from d$_{xy}$ orbital, and the $\gamma$ band is assigned as from d$_{x^2-y^2}$ orbital\cite{FChen}. Direct comparison between measurements and band structure calculations indicates a strong orbital-selective renormalization in the normal state\cite{ATamai}. Such strong renormalizaiton is consistent with the recent theoretical calculations to show that the ¡®¡¯11¡° system exhibits a stronger correlation effect in the iron-based superconductors\cite{ZPYin}.

In the optimally-doped FeTe$_{0.55}$Se$_{0.45}$ superconductor, the measured superconducting gap is nearly isotropic both near the zone center and the zone corner (Fig. 6g)\cite{HMiaoFeTeSe}. Further high-resolution measurements indicate that the superconducting gap around the Fermi surface near $\Gamma$ is anisotropic which can be represented by cos(4$\varphi$) modulation (Fig.6h)\cite{KOkazakiFeTeSe}. These ARPES results are consistent with the STM measurement on Fe(Te,Se) that points to an unconventional s-wave (S$_{\pm}$-wave symmetry) superconducting gap\cite{HanaguriFeTeSe}.  In Fe(Se,Te) superconductors, like in FeSe superconductor case\cite{SKasahara}, the Fermi energy is comparable to the superconducting gap; this system thus provides an opportunity to explore the  BCS-BEC crossover phenomenon in a superconducting material\cite{YLubashevsky,KOkazakiSRep}.

\subsection{Intercalated FeSe system}

The attempt to intercalate FeSe with potassium (K) resulted in a successful preparation of the K$_x$Fe$_{2-y}$Se$_2$ superconductor with a T$_c$ higher than 30 K\cite{JGuo}. Later on, it was found that FeSe can be intercalated with many other intercalates including Cs, Rb, Tl, Li, Na, Sr, Ca, Yb, Eu and even molecules\cite{MFang,AKMaziopaAFeSe,CHLi,HDWangAFeSe,XLChenSR,MBLucas}. This system is complicated with the coexistence of different phases. For reviews on this topic, one may refer to Refs.\cite{DXMouReview,Dagotto}.


For the completeness of the review, here we present the main ARPES results on the A$_x$Fe$_{2-y}$Se$_2$ (A=K, Cs, (Tl,K), (Tl, Rb)) system\cite{DLFengIFS,TQianAFeSe,DXMouPRL,LZhaoAFeSe,DXMouReview,XWangAFeSe,FChenAFeSe,MXu,XPWangG}. Although the intercalates can vary, the electronic structure of the superconducting phase in the A$_x$Fe$_{2-y}$Se$_2$ system turned out to be similar.  As shown in Fig. 7, the measured Fermi surface consists of two electron-like Fermi surface sheets ($\alpha$ and $\beta$ in Fig. 7g) around $\Gamma$ and one electron pocket around M ($\gamma$ in Fig. 7g) that includes two nearly degenerate Fermi surface sheets.  The bands around $\Gamma$ (Fig. 7a and 7b) is dominated by two electron bands $\alpha$ and $\beta$: the bottom of the $\alpha$ band just touches the Fermi level while the origin of the  $\beta$ band remains unclear\cite{DXMouPRL}. Around the M point, a clear electron band is observed in the vicinity of Fermi level and a hole-like band with its top about 100 meV below the Fermi level (Fig. 7c and 7d). More detailed measurements indicate that there are two electron bands with similar Fermi momenta but different band bottoms around M (Fig. 7e and f)\cite{DXMouReview}.

It was proposed that the interband scattering between the hole-like bands near $\Gamma$ and electron-like bands near M gives rise to electron pairing and superconductivity in the iron-based superconductors\cite{Kuroki,Nesting,FWangPRL,AVChubukov,VStanev,FWangEPL,IIMazinNP}. The distinct Fermi surface topology in the A$_x$Fe$_{2-y}$Se$_2$ superconductor presents a serious challenge to this picture since there is no hole-like Fermi surface present near the $\Gamma$ point.   An alternative scattering mechanism between the electron pockets was proposed\cite{Kuroki,TAMailer,FWangEEScattering,TDas}. In this case, the inter-electron pocket scattering tends to produce  d$_{x^2-y^2}$ gap symmetry. Nearly isotropic superconducting gap is observed around the $\gamma$ electron pocket at the M point (Fig. 7h)\cite{DLFengIFS,TQianAFeSe,DXMouPRL,LZhaoAFeSe,DXMouReview}. In particular, the electron Fermi pockets around the zone center provide a good opportunity to distinguish among various gap symmetries. Nearly isotropic superconducting gap is observed on the small $\alpha$ electron pocket\cite{MXu,XPWangG} and nearly isotropic superconducting gap was also observed around the larger $\beta$ electron-pocket around $\Gamma$\cite{DXMouReview}. These results are not consistent with the d-wave gap symmetry in the A$_x$Fe$_{2-y}$Se$_2$  system with only electron Fermi surface.


\section{Electronic Structure and Superconductivity of F\MakeLowercase{e}S\MakeLowercase{e} Films}


\subsection{Electronic structure and superconductivity of FeSe films grown on graphene substrate}

High-quality crystalline FeSe films can be grown on graphene substrate by a molecular beam epitaxy (MBE) method\cite{CLSong}. The graphene substrate used in this case is double-layer graphene formed on the SiC(0001) surface and the FeSe films grow as islands on the graphene layer. The grown FeSe films are along the (001) direction. One unit cell consists of one Fe layer and two Se layers (one above the Fe layer and the other below) (see Fig. 1) (loosely speaking, we will call this one unit cell of FeSe as single-layer in the following). The interaction between the FeSe film and the underlying graphene substrate is very weak so the lattice constant of the FeSe film is the same as the bulk FeSe (3.76 $\AA$). Also because of this weak interaction, the FeSe films form domains with different orientations within the (001) plane. In this case, several STM works have been done on the FeSe/graphene films\cite{CLSongScience,CLSongBoundary,CLSongMode}, but no ARPES measurement can be done due to the multi-domain nature of the films.

The superconductivity of the FeSe/graphene films behaves in a usual manner, i.e., its superconducting transition temperature decreases when the films get thinner (Fig. 8)\cite{CLSong}.  Fig. 8a shows a series of normalized tunneling conductance spectra on an 8-layer FeSe/graphene film. Clear coherence peak associated with superconductivity is observed and the gap size ($\Delta$) can be obtained by measuring the distance between the two peaks (2$\Delta$). It is 2.1 meV at 3 K for the 8-layer FeSe/grahene film with a corresponding superconducting temperature (T$_c$) at 8 K.  In comparison, bilayer FeSe/graphene film shows a lower T$_c$ that is only 3.7 K (Fig. 8b). The superconducting gap and T$_c$ are determined for the FeSe/graphene films with different layers, as shown in Fig. 8c which plots the variation of T$_c$ with the film thickness.  T$_c$ is found to be linearly proportional to the inverse of the film thickness 1/d. This is consistent with the usual expectation that T$_c$ of the ultra-thin films decreases with the decrease of the film thickness: T$_c$(d)=T$_{c0}$ (1-d/d$_c$) with T$_{c0}$ being the bulk transition temperature and d$_c$ being the critical thickness where superconductivity emerges\cite{JSimonin}. Similar behaviors are also observed in ultra-thin Pb\cite{UltraPb} and ultra-thin YBa$_2$Cu$_3$O$_{7-\delta}$\cite{YBCOTc} films. Note that no superconductivity is observed in single-layer FeSe/graphene film above 2.2 K. The extrapolated T$_{c0}$ from the layer-dependence of T$_c$ is $\sim$9.3 K that is consistent with the T$_c$ of the bulk FeSe superconductor\cite{CLSong}.

In an atomically flat and defect-free (001) surface of the FeSe/graphene film (about 30 unit cell thick) with large terraces (Fig. 9a), the topmost layer is Se-terminated (Fig. 9b)\cite{CLSongScience}.  The scanning tunneling spectroscopy (STS) probes the quasiparticle density of states and measures the superconducting gap at the Fermi energy (E$_F$) (Fig. 9c).  The maximum of the superconducting gap is measured as $\Delta$$_0$ = 2.2 meV. The most striking feature of the spectra at 0.4 K is the V-shaped dI/dV curve and the linear dependence of the quasiparticle density of states on energy near E$_F$. This feature is similar to that found in the d-wave cuprate superconductors with gap nodes\cite{OFisher} and in contrast to the U-shaped dI/dV curve observed in bulk Fe(Se,Te) superconductor\cite{HanaguriFeTeSe}. This V-shaped feature in the FeSe/graphene film explicitly indicates the existence of line nodes in the superconducting gap function\cite{CLSongScience}. Further Fermi surface topology examination (Fig. 9d) suggests that the line nodes may be related to the component of the extended s$_{\pm}$-wave form $\Delta_2$(coskx + cosky) that gives rise to nodes near M points.  The reason why the Cooper pairing is nodal in FeSe/graphene films but nodeless in bulk Fe(Se,Te) remains a theoretical challenge although it is suggested that the pnictogen height may introduce a switch between the nodal and nodeless  pairings\cite{Kurokipnictogen}.

In the FeSe/graphene films, the electron pairing with twofold symmetry, instead of four-fold symmetry, was demonstrated by direct imaging of quasiparticle excitations in the vicinity of magnetic vortex cores, Fe adatoms, and Se vacancies\cite{CLSongScience}. The twofold pairing symmetry was further supported by the observation of striped electronic nanostructures in the slightly Se-doped samples. The anisotropy was explained in terms of the orbital-dependent reconstruction of electronic structure in FeSe.  It was also discovered that, in the FeSe/graphene films, superconductivity near the twin boundaries is suppressed, which also provides a measure of the superconducting coherence length on the order of 5.1 nm\cite{CLSongBoundary}. Such a superconductivity suppression is linked to the increased Se height in the vicinity of twin boundaries\cite{CLSongBoundary}. STM/STS also revealed signatures of a bosonic mode in the local quasiparticle density of states of the superconducting FeSe/graphene films\cite{CLSongMode}.


\subsection{Electronic structure and superconductivity of FeSe films grown on SrTiO$_3$ and related substrates}









The FeSe films grown on the SrTiO$_3$ substrate exhibit surprising and different behaviors from those grown on graphene substrate\cite{LLWangReview}. When a single-layer FeSe film was grown on SrTiO$_3$ (001) substrate by a MBE method (Fig. 10), four pronounced peaks were identified at $\pm$20.1 mV and $\pm$9 mV in the tunneling spectrum at 4.2 K that are symmetric with respect to the Fermi level (Fig. 10b)\cite{QYWangCPL}. The gap at 20.1 meV, when taken as a superconducting gap, would correspond to a superconducting transition temperature near 80 K if one assumes the similar ratio between the superconducting gap and T$_c$ as in the FeSe/graphene films\cite{CLSongScience}. The temperature dependence of the gap, the observation of magnetic field-induced vortex state, and the transport measurement all indicate that the single-layer FeSe/SrTiO$_3$ film is likely supercondcting\cite{CLSongScience}. This pioneering work has generated a great excitement in the community and many experimental and theoretical works follow. On the other hand, it was found that the two-layer FeSe/SrTiO$_3$ films, grown under the same condition, show a semiconducting behavior (Fig. 10d)\cite{QYWangCPL}.


The immediate ARPES measurements on a superconducting single-layer FeSe/SrTiO$_3$ film reveal a simple and distinct electronic structure\cite{DFLiu}. The observed Fermi surface (Fig. 11a) consists of only electron pocket near the zone corner M without any Fermi surface crossing near the zone center $\Gamma$. This Fermi surface topology is the simplest among all the observed Fermi surface in the iron-based superconductors (Fig. 11); it is also dramatically different from the calculated Fermi surface of the bulk FeSe (Fig. 11d). Considering two electron-pockets around M point due to two-Fe-sites in a unit cell\cite{RPengPRL,JJLee}, the carrier concentration of this particular sample can be estimated by the area of the electron-pocket; it corresponds to $\sim$0.10 electrons per Fe.  The band structure near the zone center (left panel of Fig. 11e) is dominated by a hole-like band with its top about 80 meV below the Fermi level. The band structure across the zone corner (M point) consists of an electron-like band with its bottom about 60 meV below the Fermi level. An effective mass of $\sim$3m$_e$ can be estimated for this particular electron-like band\cite{DFLiu}.

ARPES also provides an alternative approach to examine whether the single-layer FeSe/SrTiO$_3$ is superconducting or not and its superconducting transition temperature by directly measuring the possible superconducting gap opening and its temperature dependence\cite{DFLiu}. The temperature-dependent photoemission spectra are shown in Fig. 12a; sharp peaks develop at low temperature and become gradually broadened with increasing temperature. The symmetrized photoemission spectra  is shown in Fig.12b in order to visually inspect possible gap opening and remove the effect of Fermi distribution function near the Fermi level.  There is a clear gap opening at low temperature. The gap size decreases with increasing temperature and  closes at 50$\sim$55 K.  The variation of the gap size as a function of temperature (Fig.12e) follows a standard BCS form. Momentum dependent measurements (Fig.12c and d) indicate that the gap is nearly isotropic without any sign of zero gap (node) around the Fermi surface. Since the single-layer FeSe film is highly two-dimensional, it avoids complications from three-dimensional Fermi surface. In this case, it is definitive to claim  that the gap along the electron-like Fermi surface near M is s-wave like without nodes. The near-E$_F$ band back-bending at low temperature (Fig. 12g, right panel), the sharp coherence peaks at low temperature (Fig. 12a), the BCS-form-like temperature dependence of the gap (Fig. 12e), and the momentum dependence of the gap (Fig. 12c) provide strong evidence to indicate that the gap observed here is a superconducting gap.


In the electronic phase diagram of both the high temperature cuprate superconductors and the iron-based superconductors\cite{PALee,JPaglione}, it is clear that the physical properties depend strongly on the carrier concentration, and superconductivity can be optimized at a particular carrier concentration. It is tempting to ask whether one can establish a similar phase diagram for the single-layer FeSe/SrTiO$_3$ system, and whether it is possible to optimize its superconductivity. Such an idea is motivated by the MBE growth process where the single-layer FeSe/SrTiO$_3$ film is first grown at a relatively low temperature which is non-superconducting; then it becomes superconducting only after the film is annealed in vacuum at a relatively high temperature\cite{QYWangCPL,WHZhangPRB,ZLiSTM}. To keep track on how the evolution from the as-grown non-superconducting single-layer FeSe/SrTiO$_3$ film to the vacuum-annealed superconducting film occurs, extensive ARPES measurements were carried out on the as-grown single-layer FeSe/SrTiO$_3$ film that was vacuum-annealed {\it in situ} by many sequences at different temperatures and different times\cite{SLHe}. The main results are summarized in Fig. 13. For the as-grown non-superconducting single-layer FeSe/SrTiO$_3$ film, its electronic structure (Fig. 13(e-h)) is distinct from that of the superconducting films (Fig. 13(a-d)). In this case, the underlying Fermi surface shows four strong spots around the M point (Fig. 13e). A hole-like band is present near the $\Gamma$ point but its top is closer to the Fermi level compared to that in the superconducting film. A pronounced hole-like band is observed along the $\Gamma$-M direction near the M point that is totally different from the electron-like band in the superconducting film.  The distinct electronic structures between the as-grown non-supercoucting films and the superconducting films indicate there are two different phases present in the single-layer FeSe/SrTiO$_3$ films (Fig. 13(a-h)), for convenience, the phase with the electronic structure similar to the as-grown films is called ``N"-phase while the one with the electronic structure similar to the superconducting FeSe films is called ``S"-phase\cite{SLHe}.  It turned out that the evolution proceeds in three stages during the vacuum-annealing process.\cite{SLHe}. For the as-grown non-superconducting film, after initial first-stage mild annealing, it stays in the pure N-phase (left pink region in Fig. 13i).  Further annealing leads to the emergence of the S phase; the S phase increases with the annealing process at the expense of the N phase. This gives rise to the second stage of the mixed phase region (middle light blue region in Fig. 13i). After extensive annealing at sufficiently high temperature and long time, the film can convert to pure S phase in the final third stage (right bright green region in Fig. 13i).

The annealing process also made it possible to tune the superconductivity of the single-layer FeSe/SrTiO$_3$ film by varying the carrier concentration\cite{SLHe}. The carrier concentration of the S phase during the annealing process can be obtained by calculating the area of the electron-like Fermi surface near M.  It has been shown that during the vacuum annealing process, the S phase gets more and more electron-doped, with the electron concentration increasing from 0.07 to 0.12 electrons per Fe\cite{SLHe}. The corresponding gap measurements indicate that the gap size and the gap closing temperature increase with the annealing process (Fig. 13i). A gap of $\sim$19 meV and a gap closing temperature of $\sim$ 65 K were achieved for the annealed single-layer FeSe/SrTiO$_3$ film (Fig. 13i)\cite{SLHe}. This provides first electronic evidence of superconductivity at $\sim$65 K in the single-layer FeSe/SrTiO$_3$ films that was also confirmed by other groups\cite{STan,JJLee}. Such a gap increase with vacuum annealing was also observed in the STM/STS measurements\cite{WHZhangPRB}.

Since the S phase appears in the early stage of vacuum annealing of the single-layer FeSe/SrTiO$_3$ films which coexists with the N phase\cite{SLHe}, it is intriguing to ask whether the S phase becomes superconducting as long as it appears. To address this issue, the evolution of the electronic structure and the energy gap as a function of the carrier concentration is studied for the S phase of the single-layer FeSe films\cite{JFHe}. It was found that, at low temperature, there is a gap opening along the underlying electron-like Fermi surface near M at low carrier concentration, as seen from the symmetried photoemission spectra around the Fermi surface (Fig. 14b). As the carrier concentration of the S phase increases, the gap size decreases and reaches zero at a critical carrier concentration around 0.089 electrons/Fe (Fig. 14a and 14b). Further increase of the carrier concentration results in another gap opening at low temperature. From the appearance of the coherence peak and the gap closing with increasing temperature (Fig. 14d), this high carrier concentration gap is apparently the superconducting gap we have discussed before. For the low carrier concentration gap, it shows quite different behaviors from the high carrier concentration superconducting gap.  The broad photoemission peaks, weak temperature dependence of the gap (Fig. 14c) and its relatively large size compared to the superconducting gap are similar to those observed in the insulating heavily-underdoped cuprates\cite{pyy}. We believe the low carrier concentration energy gap is consistent with an insulating gap. Therefore, there  is an insulator-superconductor transition across the critical carrier concentration of 0.089 in the S phase of the single-layer FeSe/SrTiO$_3$ films\cite{JFHe}. It was noted that such a transition exhibits many similar behaviors as that found in the heavily underdopd cuprates\cite{pyy}. At present, the origin of this insulator-superconductor transition remained to be understood which may be associated with the two-dimensionality that enhances electron localization or correlation.  In a multi-orbital system like the iron-based compounds, the carrier density-induced Mott transition may be realized in an orbital-selective fashion\cite{RYuOrbitalS}.  Such an orbital-selective Mott transition has been examined in A$_x$Fe$_{2-y}$Se$_2$ superconductor\cite{MYi}. In the S phase of the single-layer FeSe/SrTiO$_3$ films, it is possible that the same mechanism is at operation for the observed carrier concentration-induced insulator-superconductor transition\cite{JFHe}.

The electronic structure and superconductivity of the FeSe/SrTiO$_3$ films show drastically different layer-dependence from that of the FeSe/graphene films (Fig. 8)\cite{CLSong}, as well as FeSe films grown on the MgO substrate\cite{RSchneider}. For the FeSe/SrTiO$_3$ films, while single-layer FeSe/SrTiO$_3$ film can become superconducting, STM/STS does not find signature of superconductivity in the second and multiple-layer FeSe/SrTiO$_3$ films regardless of how the samples are annealed\cite{QYWangCPL,ZLiSTM}.  Such a dramatic difference between the single-layer, double-layer and multiple-layer FeSe/SrTiO$_3$ films is surprising if one considers the usual layer-dependence found in the FeSe/graphene films\cite{CLSong}. To address this issue, a comparative investigation between the single-layer and double-layer FeSe/SrTiO$_3$ films is carried out by performing a systematic angle-resolved photoemission study on the samples annealed in vacuum\cite{XLiu}.  The as-prepared double-layer FeSe/SrTiO$_3$ film exhibits electronic structure that is similar to the N phase of the single-layer FeSe/SrTiO$_3$ film (Fig. 15).  The double-layer FeSe/SrTiO$_3$ film shows a quite different doping behavior from the single-layer film. It is hard to get doped and remains in the semiconducting/insulating state even under an extensive annealing condition. However, after sufficient vacuum annealing, the double-layer FeSe/SrTiO$_3$ film may follow similar doping trend as in the single-layer film although superconductivity has not been realized.  Such a behaviour is understood as originating from the much reduced doping efficiency in the bottom FeSe layer of the double-layer FeSe/SrTiO$_3$ film from the FeSe-SrTiO$_3$ interface when the dominant doping process comes from the electron transfer from the SrTiO$_3$ surface\cite{XLiu}.

Detailed ARPES study on the layer-dependence of the electronic structure for the FeSe/SrTiO$_3$ films have been reported in \cite{STan,JJLee} and typical results are shown in Fig. 16\cite{STan}. It is found that the electronic structure of the FeSe/SrTiO$_3$ films becomes similar when the number of layers is above two. In this case, the Fermi surface is quite different from the S phase of the superconducting single-layer FeSe/SrTiO$_3$ film in that: the Fermi surface of the multi-layer FeSe/SrTiO$_3$ films  consists of four strong spots near the M point, and there are bands crossing the Fermi level around the $\Gamma$ point (Fig. 16a). It was also found that the multi-layer FeSe/SrTiO$_3$ films experience a transition at a high temperature where the electronic structure exhibits a change above and below the transition temperature\cite{STan}. The transition temperature increases as the number of FeSe layers decreases (Fig. 16b). Such a transition is attributed to the spin-density-wave formation and the layer-dependent transition temperature is associated with the strain difference in samples with different layers\cite{STan}.

We note that the N phase of the single-layer FeSe/SrTiO$_3$ film is not observed in \cite{STan,JJLee}. Also there are discrepancies on the electronic structure of the double-layer FeSe/SrTiO$_3$ films between different groups\cite{XLiu,STan,JJLee}. The difference can be attributed to the different preparation conditions used and possible mixing of the single-layer and double-layer FeSe/SrTiO$_3$ films. In our case, we have clearly observed pure N phase in both the single-layer and double-layer FeSe/SrTiO$_3$ films\cite{SLHe,XLiu}. In fact, the electronic structure of the N phase in the single-layer and double-layer FeSe/SrTiO$_3$ films is rather similar to that found in multi-layer films\cite{STan,JJLee} except for some qualitative band position difference. The electronic structure of the N phase in the single-layer and double-layer FeSe/SrTiO$_3$ films also experience a transition at high temperature (Fig. 17), the transition temperature in single-layer FeSe/SrTiO$_3$ is the highest among all the FeSe/SrTiO$_3$ films and basically follows the trend found for multi-layer FeSe/SrTiO$_3$ films\cite{STan}. Such a transition is reminiscent of the electronic structure change in the parent compound BaFe$_2$As$_2$ across the structural/magnetic transition temperature ($\sim$140 K for BaFe$_2$As$_2$)(Fig. 17e and 17f)\cite{GDLiu}.  In bulk FeSe, there is a structural transition near 90 K\cite{TMMcQueen}. Recently, an electronic structure change is also reported in the ARPES stiudy of the bulk FeSe crystals across this structural transtion\cite{KNakayama,TShimojima}. Since there is no static magnetic transition found in bulk FeSe, whether the electronic structure transition in the FeSe/SrTiO$_3$ films is caused by spin-density-wave formation needs further investigations and direct magnetic measurements.

The dramatic difference of superconductivity between the FeSe/SrTiO$_3$ films and the FeSe/graphene films points to the key role played by the growth substrate. Furthermore, as the superconductivity of the FeSe superconductor is found to be sensitive to pressure in the bulk form\cite{SMedvedev} or strain in the thin film form\cite{YFNieFeSe,FNebashima}, it is natural to ask whether the strain may play the key role in enhancing superconductivity in the FeSe/SrTiO$_3$ films. As the bulk FeSe has a lattice constant of 3.76 $\AA$ and SrTiO$_3$ has a lattice constant of 3.90 $\AA$, when FeSe films are epitaxially grown on the SrTiO$_3$ substrate, the FeSe films will experience a tensile stress from the underlying SrTiO$_3$ substarte\cite{QYWangCPL}.  To check on the effect of strain on the superconductivity of the FeSe films, different substrates with disparate lattice constants have been used to grow FeSe films\cite{RPengPRL,RPengNC}.  With the substrate varying from SrTiO$_3$ to SrTiO$_3$/KTaO$_3$ (SrTiO$_3$ buffer-layers grown on KTaO$_3$ substrate as the substrate for FeSe films), to BaTiO$_3$/KTaO$_3$ (BaTiO$_3$ buffer layers grown on KTaO$_3$ substrate as the substrate for FeSe films), there are changes on the corresponding electronic structures of the FeSe films (Fig. 18a and 18b), but the gap closing temperature shows little change within the range of (70$\pm$5) K\cite{RPengPRL,RPengNC}. This seems to indicate that the tensile stress on the FeSe films exerted from the underlying substrate does not affect superconductivity of the FeSe films significantly.


The latest episode sees the observation of replica band in the band structure of the superconducting single-layer FeSe/SrTiO$_3$ films (Fig. 19)\cite{JJLee}. As shown in Fig. 19, the main hole-like band near $\Gamma$ point (band d in Fig. 19b), the electron-like band (band a in Fig. 19c) and hole-like band (band b in Fig. 19c) all produce weak ``replica" bands that are $\sim$100 meV below the original bands. Such  replica bands persist above the gap opening superconducting temperature (Fig. 19d-g). The replica bands appear weak in other measurements\cite{DFLiu,SLHe,STan} possibly because of the photoemission matrix element effect causes by different measurement conditions.  These replica bands have been understood as due to the presence of bosonic modes, most probably optical phonons in SrTiO$_3$, that couple to the FeSe electrons with only a small momentum transfer\cite{JJLee}. It was also suggested that such interfacial coupling could assist superconductivity in most channels and this coupling is responsible for the T$_c$ enhancement in the single-layer FeSe/SrTiO$_3$ films\cite{JJLee}.


\subsection{Transport and magnetic measurements of FeSe/SrTiO$_3$ films}

Although ARPES and STM/STS have observed gap opening and the behaviors of the gap are consistent with the superconducting gap such as the sharp coherence peak, the BCS form of the gap size as a function of temperature, and the particle-hole symmetry around the Fermi surface\cite{QYWangCPL,DFLiu,SLHe,STan}, direct evidence from the transport and magnetic measurements is necessary to establish whether the FeSe/SrTiO$_3$ films are truly superconducting or not. Such measurements on ultra-thin films are challenging because of the complication of the sample deterioration to the air, SrTiO$_3$ substrate conductance and very weak magnetic signal. Initial transport measurement on a five-layer FeSe/SrTiO$_3$ film covered by 20-nm-thick amorphous silicon protection layer saw a transition with an onset temperature of $\sim$53 K (Fig. 20a)\cite{QYWangCPL}.  Later on, the single-layer FeSe/SrTiO$_3$ film was better protected by covering a 10-layer FeTe film followed by a 30-nm-thick amorphous silicon layer so that the sample can be stable at ambient condition for a relatively long time. The resistance as a function of temperature clearly reveals the appearance of superconductivity: the resistance begins to drop at 54.5 K and reaches zero at 23.5 K with an onset T$_c$ of 40.2K (Fig. 20b)\cite{WHZhangCPL}.  Magnetic measurement on the same sample reveals a diamagnetic response with an abrupt change of both in-phase and out-of-phase signal at 21 K (Fig. 20d) that is consistent with the zero resistance critical temperature from the transport measurement (Fig. 20b). This has unambiguously demonstrated that the single-layer FeSe/SrTiO$_3$ film under measurement is superconducting although the measured T$_c$ is not as high as that expected from the STM/ARPES measurements. This is understandable because  it is known that the superconducting transition temperature of the FeSe/SrTiO$_3$ films is sensitive to the carrier concentration\cite{SLHe}.  Also the protection layer may affect the superconductivity of the single-layer FeSe/SrTiO$_3$ film, in particular, FeTe is known to be magnetic at low temperature that may suppress the superconductivity of the adjacent FeSe film.  Superconductivity is also reported in double-layer FeSe/SrTiO$_3$ films from the transport measurements\cite{YSun}.  Hints of higher transition temperatures are spotted in some FeSe/SrTiO$_3$ films\cite{LZDeng} that need further investigations.

It is ideal to carry out {\it in situ} transport and magnetic measurements on the FeSe/SrTiO$_3$ films in order to remove the effect of sample damage and protection layers. The {\it in situ} transport measurements have been performed lately on the single-layer FeSe/SrTiO$_3$ film\cite{JFGe}. Surprisingly, the measured zero resistance temperature reaches as high as 109 K (Fig. 20c) that is well above the T$_c$$\sim$80 K expected from the 20.1 meV gap from the STM/STS measurements\cite{QYWangCPL} and the T$_c$$\sim$65 K from ARPES measurements\cite{SLHe,STan}. Further measurements are needed to reproduce and confirm this exciting result.

\section{Implications and theoretical understanding}

\subsection{Implications}


The discovery of high temperature superconductivity in the single-layer FeSe/SrTiO$_3$ films has important implications in understanding the superconductivity mechanism and the pairing symmetry in the iron-based superconductors. In most of the iron-based superconductors, the Fermi surface consists of hole-like Fermi surface sheets near the Brillouin zone center ($\Gamma$ point) and electron-like Fermi surface sheets near the zone corner M($\pi$,$\pi$) point (Fig. 2)\cite{DJSingh}. It has been proposed that the electron scattering between the hole-like bands around $\Gamma$ and electron-like bands near M is responsible for electron pairing and superconductivity of the iron-based superconductors has a dominant s$\pm$ superconducting order parameter\cite{Kuroki,Nesting,FWangPRL,AVChubukov,VStanev,FWangEPL,IIMazinNP}. The absence of Fermi surface near $\Gamma$ in the superconducting single-layer FeSe/SrTiO$_3$ film\cite{DFLiu}, together with that only electron-like Fermi surface sheets are observed in the intercalated A$_x$Fe$_{2-y}$Se$_2$ superconductors\cite{DLFengIFS,TQianAFeSe,DXMouPRL},  rules out the necessity of the electron scattering between the hole-like bands near $\Gamma$ and electron-like bands near M in this ``Fermi nesting" picture.

The observation of the nearly isotropic superconducting gap around the electron-like Fermi surface near M point (Fig. 12c)\cite{DFLiu} also provides crucial information in understanding the pairing symmetry of the iron-based superconductors. With the absence of hole-like Fermi surface near $\Gamma$, electrons can only scatter across the Fermi surface sheets between M points which is predicted to result in a d-wave superconducting gap\cite{Kuroki,TAMailer,FWangEEScattering,TDas}.  The d$_{xy}$-like superconducting order parameter is obviously ruled out because it will produce zero gap on the electron-like Fermi surface near M. With an alternative d$_{x^2-y^2}$-like superconducting order parameter that can avoid the direct gap nodes, it is argued that two Fermi surface sheets around a given M point with opposite phases can also give rise to gap nodes around the Fermi surface\cite{IMazinNode}. In this case, it remains not consistent with the observation of nodeless gap on the electron-like Fermi surface near M point.  The superconducting pairing symmetry in the single-layer FeSe/SrTiO$_3$ films and the A$_x$Fe$_{2-y}$Se$_2$ system remains puzzling that asks for further experimental and theoretical efforts.

The different behaviors between the FeSe/graphene and FeSe/SrTiO$_3$ films clearly indicate the critical role played by the interface in giving rise to the high temperature superconductivity.  The discovery of high temperature superconductivity in the single-layer FeSe/SrTiO$_3$ films represents the first clear case of interface-enhanced superconductivity. Interface (and surface) superconductivity has a long history which is expected to increase superconducting temperature\cite{Ginzburg}. Superconductivity has been reported in various interface systems, including those between two insulators\cite{NReyren}, between one insulator and one metal\cite{AGozar}, and between one metal and one semiconductor\cite{TZhang}. However, in all these cases, the observed T$_c$ in the interface is not obviously higher than the maximum T$_c$ of its individual constituents. In the single-layer FeSe/SrTiO$_3$ films, its superconducting transition temperature is much higher than the maximum T$_c$ of either FeSe system or SrTiO$_3$ system, thus establishing a convincing case of interface superconductivity. This discovery proves the concept of interface-enhancement of superconductivity and opens a door to further explore for new superconductors with higher T$_c$.

\subsection{Theoretical understanding}

The FeSe/SrTiO$_3$ films have attracted much theoretical attention, in regard to understanding its unique electronic structure, carrier doping mechanism and the possible origin of high temperature superconductivity\cite{KLiu,YYXiang,TBazhirov,FWZheng,HYCao,QQYe,JBang,TBerlijn,NNHao}. By using first-principles calculations, the atomic and electronic structures of the single-layer and double-layer FeSe/SrTiO$_3$ films were studied\cite{KLiu}. The band structure of free single-layer FeSe film shown (Fig. 21a) is similar to that of the bulk FeSe (Fig. 4)\cite{ASubedi}.  It is found that both single-layer and double-layer FeSe films on the SrTiO$_3$ substrate behave like a slightly doped semiconductor with a collinear antiferromagnetic order on the Fe ions and a Dirac cone-like band around the $\Gamma$ point (Fig. 21b)\cite{KLiu}. These results show difference from the measured results, for both the N phase and S phase of the single-layer and double-layer FeSe/SrTiO$_3$ films\cite{DFLiu,SLHe,STan,XLiu}. It was further shown that the SrTiO$_3$ substrate-induced tensile strain tends to stabilize the collinear antiferromagnetic state in FeSe thin film by enhancing of the next-nearest-neighbor super-exchange antiferromagnetic interaction bridged through the Se atoms\cite{HYCao}. The calculated Fermi surface considering the collinear antiferromagnetic order and electron-doping can be consistent with experimental observations that the hole-like Fermi surface sheets near $\Gamma$ disappear\cite{HYCao}.  In the presence of constrained magnetization, it was also shown that the observed Fermi surface can be described by the checker-board antiferromagnetic spin pattern\cite{TBazhirov,FWZheng}. In this case, it was found that the system has a considerable charge transfer from SrTiO$_3$ substrate to the FeSe layer, and so has a self-constructed electric field. The disappearance of the Fermi surface around the $\Gamma$ point can also be explained by the antiferromagnetic checkerboard phase with charge doping and electric field effects\cite{FWZheng}. These results show considerable consistency with the behavior of the N phase of the FeSe/SrTiO$_3$ film\cite{SLHe,XLiu}.  So far, most theoretical calculations involve antiferromagnetic order, and overall show some qualitative agreement with the experimental observations. However, we note that,  the ARPES results are only suggestive that the N phase of the FeSe/SrTiO$_3$ films may be magnetic\cite{SLHe} or the low temperature form of the multi-layer FeSe/SrTiO$_3$ films may be in the  spin-density-wave state\cite{STan}, further direct evidence is needed to pin down on their magnetic nature.

It is clear from ARPES measurements that the superconducting S phase of the single-layer FeSe/SrTiO$_3$ films is electron-doped\cite{DFLiu,SLHe,STan,JJLee}. The origin of the electron-doping may have two possibilities: oxygen vacancy formation on the SrTiO$_3$ surface that gives rise to electron doping, and the formation of Se vacancies in the FeSe film\cite{DFLiu}. Further ARPES results substantiate the picture of the oxygen vacancy formation on the SrTiO$_3$ surface\cite{STan,XLiu}. The role of the oxygen vacancy on the top layer of the SrTiO$_3$ substrate was also investigated by first principles calculations\cite{JBang}. It shows that the oxygen vacancies strongly bind the FeSe layer to the substrate giving rise to a (2$\times$1) reconstruction, and also serve as the source of electron doping. Theoretical investigation on the doping effect of Se vacancies in the FeSe/SrTiO$_3$ films leads to a somehow counter-intuitive result that, in terms of the Fe-3d bands, Se vacancies behave like hole dopants rather than electron dopants as usually expected\cite{TBerlijn}. Such results are considered to exclude Se vacancies as the origin of the large electron pockets measured by ARPES\cite{TBerlijn}.

The most prominent issue concerning the superconductivity of the single-layer FeSe/SrTiO$_3$ films is on the origin of its dramatic T$_c$ enhancement. The discovery of high temperature superconductivity in the single-layer FeSe/SrTiO$_3$ films was motivated by the idea that strong electron-phonon coupling can enhance superconductivity\cite{QYWangCPL}. Subsequent study on the effects of screening due to the
SrTiO$_3$ ferroelectric phonons on the Cooper pairing in FeSe shows that it can significantly enhance the energy scale of Cooper pairing and even change the pairing symmetry\cite{YYXiang}. It was shown that the role of the SrTiO$_3$ substrate in increasing the electron-phonon coupling and the resulting higher Tc is two-fold. First, the interaction of the FeSe and TiO$_2$ terminated face of SrTiO$_3$ prevents the single-layer FeSe from undergoing a shearing-type (orthorhombic) structural phase transition. Second, the substrate allows an antiferromagnetic ground state of FeSe which
opens certain electron-phonon coupling channels within the single-layer that are prevented by symmetry in the non-magnetic phase\cite{QQYe}. The replica bands observed in the superconducting single-layer FeSe/SrTiO$_3$ films provide experimental evidence on the coupling of electrons in FeSe layer with the SrTiO$_3$ phonons\cite{JJLee}. Further theoretical analysis suggests that such interfacial electron-phonon coupling assists superconductivity in most channels, and is responsible for raising the superconducting temperature in single-layer FeSe/SrTiO$_3$ films\cite{JJLee}.

\section{Summary and Perspective}








In this brief review, we have shown that the FeSe superconductor and related systems have exhibited rich and fascinating phenomena and physical properties.  Since the report of possible high temperature superconductivity in the single-layer FeSe/SrTiO$_3$ films\cite{QYWangCPL}, great progress has been made in understanding related issues both experimentally and theoretically. Some questions are answered, and more questions are asked. Since this is a burgeoning and fast-evolving field, it is natural that many issues remain to be addressed. Here we briefly list some further issues that are either particular to the FeSe/SrTiO$_3$ systems or more general to the iron-based superconductors.\\

(1). {\it Interface superconductivity?} So far many experiments point to the interface superconductivity in the single-layer FeSe/SrTiO$_3$ films. More decisive evidence is needed to pin down on the nature of the interface superconductivity. What new effects can be expected from an interface superconductivity system?\\

(2). {\it Doping mechanism?} It has been shown that during the vacuum annealing process, the single-layer FeSe/SrTiO$_3$ films can be electron-doped. It has also been proposed that oxygen vacancies on the SrTiO$_3$ surface may play the role of electron-doping\cite{DFLiu,SLHe,STan}. How to directly prove the role of oxygen vacancies in electron-doping?  Recent STM measurements on the vacuum annealed single-layer FeSe/SrTiO$_3$ film found that\cite{WHZhangPRB}, there are some extra Se adatoms on the surface of the as-prepared FeSe film and the sample is insulating. In the initial stage of the annealing process, these extra Se adatoms are gradually removed. Further annealing leads to the formation of Se vacancies. What are the role of these Se adtoms and Se vacancies on the doping of FeSe films? \\

(3). {\it Role of electron-phonon coupling?} There is experimental evidence to show the existence of electron-phonon coupling in the FeSe/SrTiO$_3$ films\cite{JJLee}. There are theoretical arguments that such electron-phonon coupling can enhance superconductivity\cite{QYWangCPL,YYXiang,QQYe,JJLee}. However, more experimental evidence is needed to prove that such an electron-phonon coupling is the cause of the T$_c$ enhancement in the single-layer FeSe/SrTiO$_3$ films.\\

(4). {\it Pairing mechanism?} The observation of the distinct Fermi surface topology in the superconducting FeSe/SrTiO$_3$ films\cite{DFLiu} has ruled out the possibility of Fermi surface nesting picture in giving rise to superconductivity in the single-layer FeSe/SrTiO$_3$ films. With only electron-pockets present near M point, what is the pairing mechanism then?\\

(5). {\it Pairing order parameter?} With only electron pockets present near M points in the single-layer FeSe/SrTiO$_3$ films\cite{DFLiu,SLHe,STan,JJLee}, it is predicted that d-wave pairing may dominate\cite{Kuroki,TAMailer,FWangEEScattering,TDas} which is not consistent with the experimental results of nearly isotropic nodeless superconducting gap\cite{DFLiu}. What can be the superconducting order parameter for the superconducting single-layer FeSe/SrTiO$_3$ films?\\

(6). {\it Theoretical understanding of electron structure.}  The unique electronic structure of the N phase and S phase of the single-layer FeSe/SrTiO$_3$ film and the multi-layer FeSe/SrTiO$_3$ films poses a challenge to band structure calculations and theoretical understandings. More efforts are needed to understand the electronic structure by including the effect of the substrate, vacancy and disorder, electron correlation, the spin-orbit coupling and so on. \\

(7). {\it Inconsistency of superconductivity measurements.} There are inconsistencies among different measurements on the superconducting properties of the FeSe/SrTiO$_3$ films. First, transport measurements on the single-layer FeSe/SrTiO$_3$ films give different T$_c$, varying from 23.5 K (zero resistance\cite{WHZhangCPL}) to 109 K\cite{JFGe}. Second, there is an obvious difference of  T$_c$ on single-layer FeSe/SrTiO$_3$ films between transport measurements\cite{QYWangCPL,WHZhangCPL,YSun,JFGe} and STM/ARPES measurements\cite{QYWangCPL,DFLiu,SLHe,STan}. Third, for multi-layer FeSe/SrTiO$_3$ films, STM/STS and ARPES indicate they are not superconducting\cite{QYWangCPL,XLiu}. However, transport measurements show sign of superconductivity in double-layer FeSe/SrTiO$_3$ film\cite{YSun},five-layer FeSe/SrTiO$_3$ film\cite{QYWangCPL}. How to reconcile these differences?\\

(8). {\it Nature of the N phase and S phase in the single-layer FeSe/SrTiO$_3$ films.}  With vacuum annealing, two different phases (N phase and S phase) with distinct electronic structures are clearly observed\cite{SLHe}. It remains unclear about the real difference between the N phase and S phase, i.e., whether the electronic structure difference is caused by the difference in crystal structure, composition or the carrier concentration. It is also not clear how the N phase and S phase distribute in a real space.\\

(9). {\it Magnetic state?} ARPES has provided suggestions that the N phase of the single-layer FeSe/SrTiO$_3$ film may be magnetic\cite{SLHe}, and the low-temperature form of the multi-layer FeSe/SrTiO$_3$ films may be in the spin-density-wave state\cite{STan}. Direct evidence is needed to pin down on their magnetic nature.\\

(10). {\it Further enhancement of T$_c$?} The superconducting transition temperature of the FeSe/SrTiO$_3$ film is sensitive to carrier concentration\cite{SLHe}. T$_c$ has not reached the maximum in the doping range studied; more electron doping may result in even higher T$_c$\cite{SLHe}. For double-layer or multi-layer FeSe/SrTiO$_3$ films, it has been found that the absence of superconductivity is related to the insufficient electron doping\cite{XLiu}. It is interesting to explore, if sufficient electron doping is provided, whether double-layer or multi-layer FeSe/SrTiO$_3$ films can become superconducting or exhibit even higher T$_c$ than the single-layer case. So far, SrTiO$_3$ and related BaTiO$_3$ substrate are proven to be magic in producing high temperature superconductivity in the single-layer FeSe films. Are there other substrates present that can produce similar effect and even higher T$_c$?\\

The surprising discovery of high temperature superconductivity in the FeSe/SrTiO$_3$ films has provided a new platform for investigating interface superconductivity, superconductivity mechanism of the iron-based superconductors, and opened a new path in exploring for new high temperature superconductors.  Theoretical calculations suggest robust topological phases may be realized in the single-layer FeSe/SrTiO$_3$ films which is intertwined with superconductivity\cite{NNHao}. The ultra-thin nature of the FeSe/SrTiO$_3$ films is also ideal for making heterostructures\cite{WLWang}; the topological insulator/superconductor heterostructure has been proposed as a prototype to detect Majorana fermions\cite{LFu}. The ideal two-dimensional character, dichotomy between single-layer and double-layer films,  and high T$_c$ superconductivity open a door for potential device fabrication and applications. We have already witnessed surprises brought up by the single-layer FeSe films, we can expect more  surprises to come in the near future.

$^{*}$Corresponding author: XJZhou@aphy.iphy.ac.cn

\vspace{3mm}

\noindent {\bf Acknowledgement} We thank Wenhao Zhang, Yun-Bo Ou, Fansen Li, Chenjia Tang, Qing-Yan Wang, Zhi Li, Lili Wang, Xi Chen, Xucun Ma, Qikun Xue, Jiangping Hu, Q. M. Si and Dunghai Lee for collaborations and helpful discussions.  XJZ thanks financial support from the NSFC (11190022,11334010 and 11374335)£¬ the MOST of China (973 program No: 2011CB921703 and 2011CBA00110) the Strategic Priority Research Program (B) of the Chinese Academy of Sciences (Grant No. XDB07020300).

\newpage

\begin{figure}[htbp]
\begin{center}
\includegraphics[width=1.0\columnwidth,angle=0]{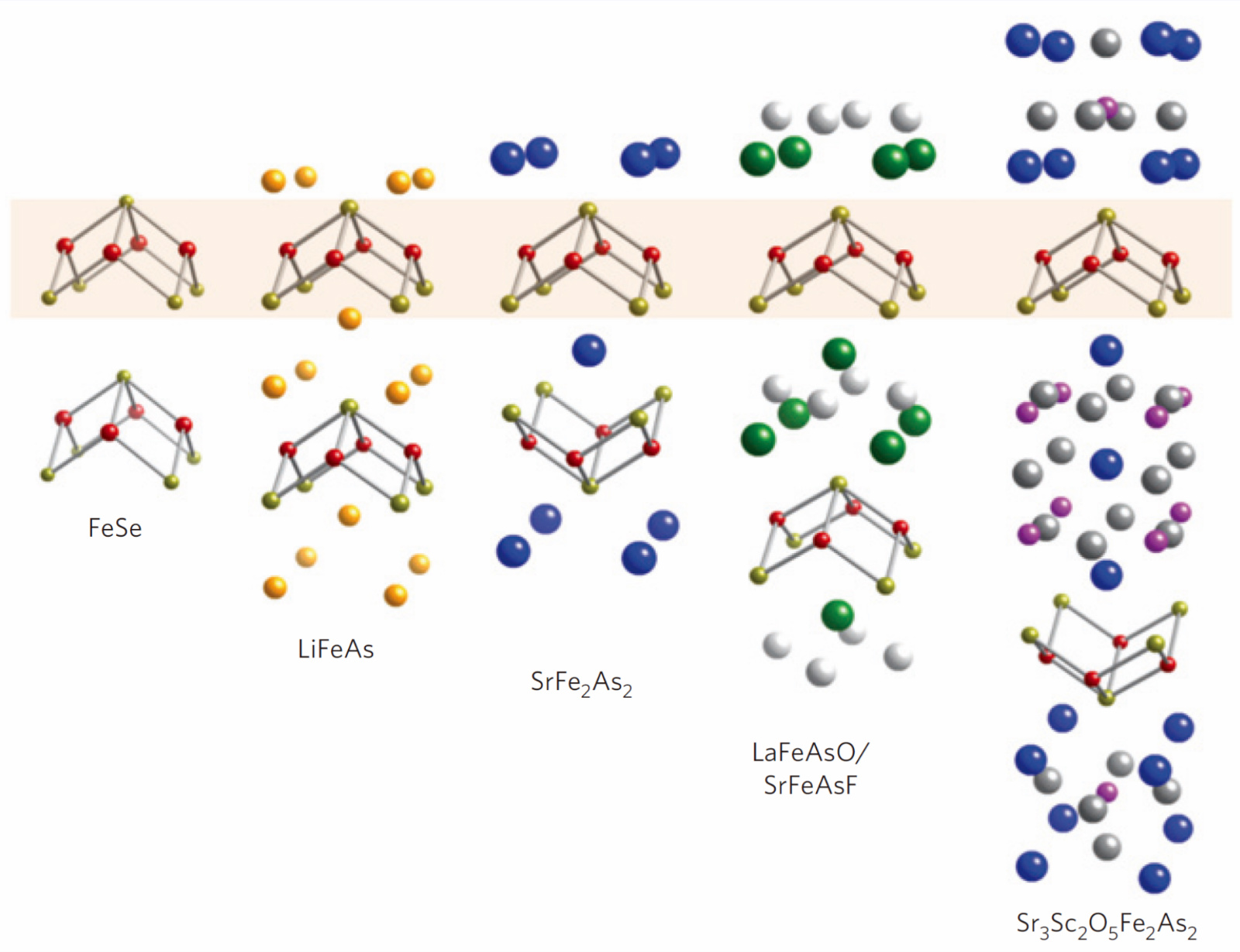}
\end{center}
\caption{\textbf{Several major classes of the iron-based superconductors.}  All of these iron-based superconductors contain FeSe or FeAs building blocks (the iron ions are shown in red and the pnictogen/chalcogen anions are shown in gold) that are essential for the occurrence of superconductivity. Reprinted from \cite{JPaglione}.}
\end{figure}

\begin{figure}[htbp]
\begin{center}
\includegraphics[width=1.0\columnwidth,angle=0]{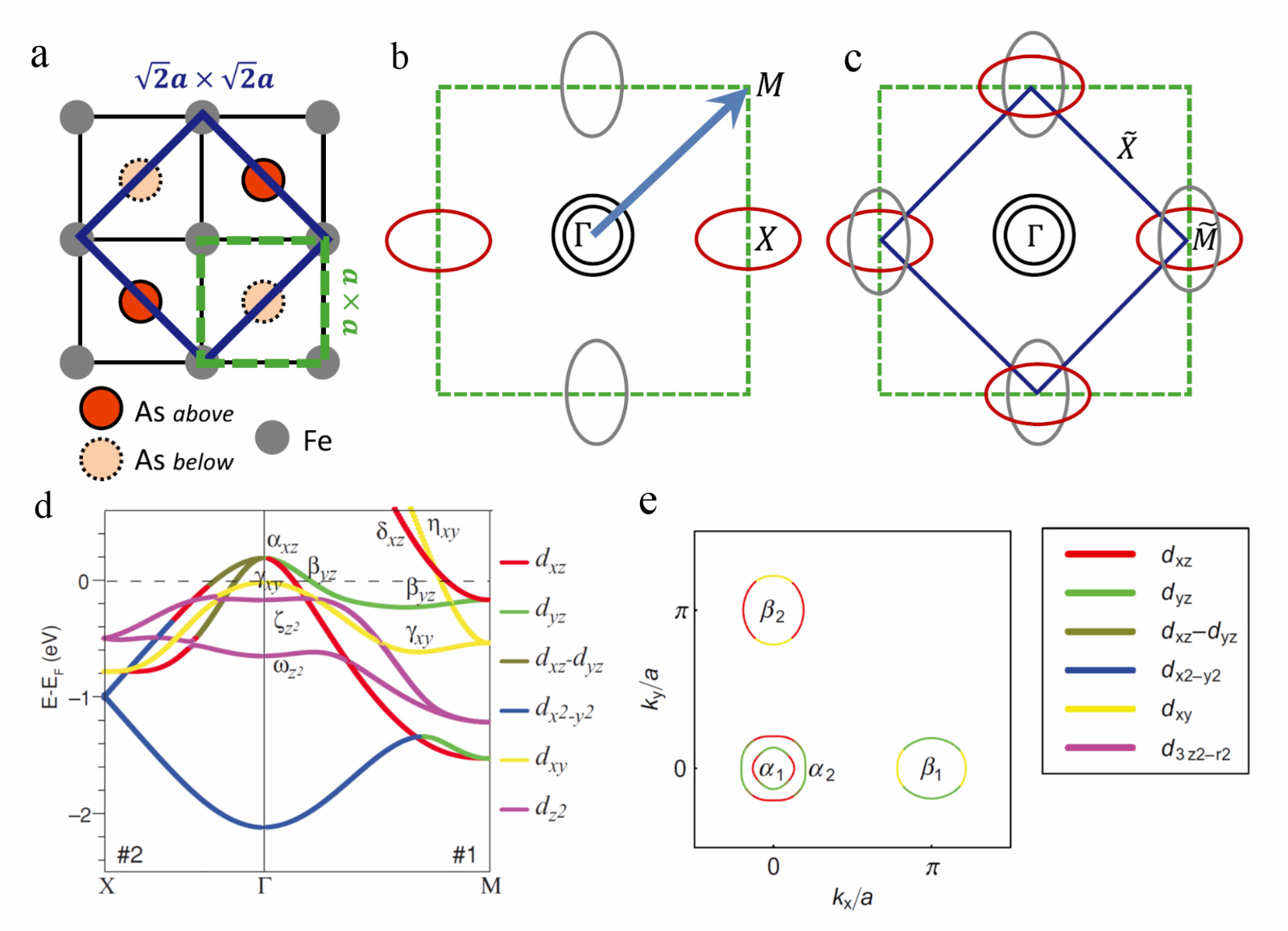}
\end{center}
\caption{\textbf{Brillouin zones of the iron-based superconductors containing one iron and two irons in a unit cell and the orbital-related electronic structure.} (a) FeAs lattice indicating As above and below the Fe plane. Dashed green and solid blue squares represent 1- and 2-Fe unit cells, respectively. (b) Schematic two-dimensional Fermi surface in the 1-Fe Brillouin zone whose boundaries are indicated by a green dashed square. The arrow indicates folding wave vector to convert from 1-Fe zone to the 2-Fe zone. (c) Fermi sheets in the folded Brillouin Zone whose boundaries are now shown by a solid blue square. In this case, the size of the first Brillouin zone is half of that of the 1-Fe case (dashed green squeare). Reprinted from \cite{Hirschfeld}. (d) A typical orbital characters of the bands in the iron pnictides as shown in Ref.\cite{SGraser,YZhangorbital}. (e) The corresponding Fermi surface sheets of the five-band model for the undoped compound in Ref.\cite{SGraser}.}
\end{figure}

\begin{figure}[htbp]
\begin{center}
\includegraphics[width=1.0\columnwidth,angle=0]{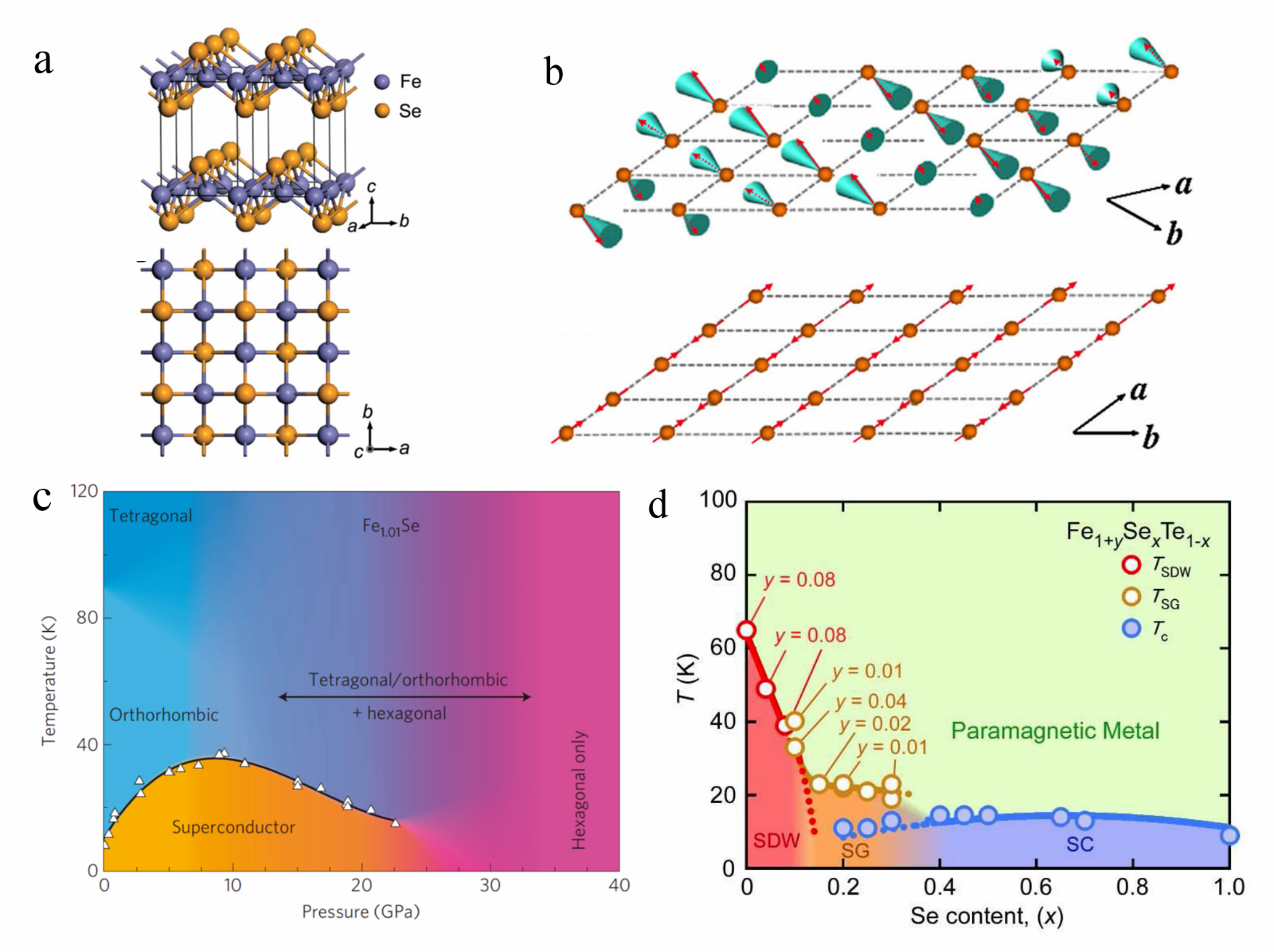}
\end{center}
\caption{\textbf{Crystal structure, magnetic structure and phase diagram of bulk Fe$_{1.01}$Se and the phase diagram of the Fe$_{1+y}$Se$_x$Te$_{1-x}$ system.} (a) Schematic crystal structure of $\beta$-FeSe. Four unit cells are shown to reveal the layered structure. Reprinted from \cite{MKWu}. (b) Magnetic structures of $\alpha$-FeTe (upper panel) and BaFe$_2$As$_2$ (lower panel), which are shown in the primitive Fe square lattice. Reprinted from \cite{WBao}. (c) Phase diagram of Fe$_{1.01}$Se as a function of pressure. Reprinted from\cite{SMedvedev}. (d) Phase diagram of Fe$_{1+y}$Se$_x$Te$_{1-x}$ with y $\sim$0 as a function of x and T.  Reprinted from \cite{Nkatayama}.}
\end{figure}

\clearpage

\begin{figure}[htbp]
\begin{center}
\includegraphics[width=1.0\columnwidth,angle=0]{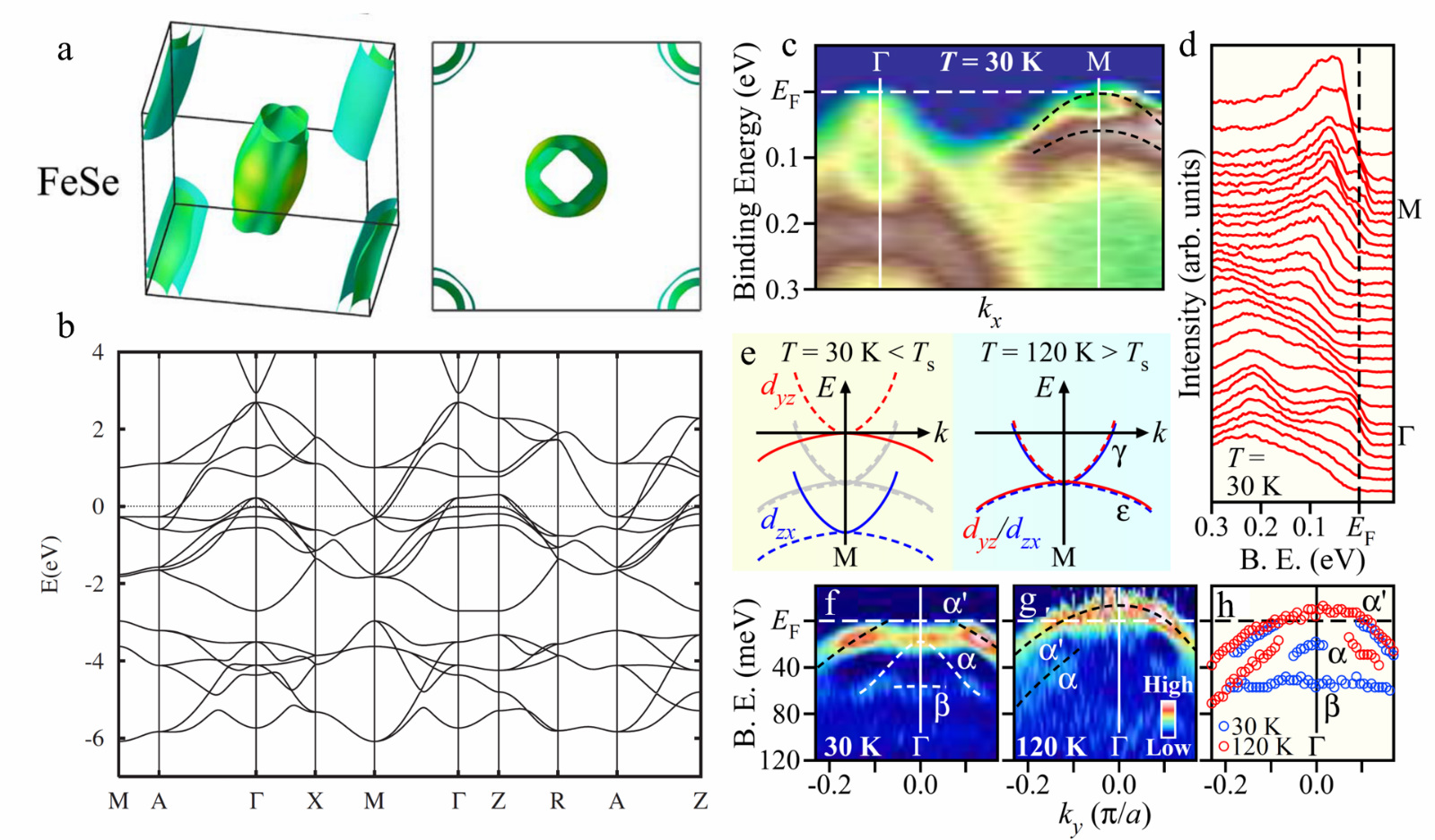}
\end{center}
\caption{\textbf{Fermi surface and band structure of the bulk FeSe from band structure calculations and angle-resolved photoemission measurements.} (a) (b) LDA calculated Fermi surface and band structure of bulk FeSe from \cite{ASubedi}. (c) (d) ARPES intensity and corresponding EDCs, respectively, along the $\Gamma$-M cut at T = 30 K. (e) Schematic band diagram around the M point below/above the structural transition temperature T$_s$. Red and blue curves indicate the d$_{yz}$ and d$_{zx}$ orbitals, respectively. Solid and dashed curves represent the band dispersion along the (0, 0)-($\pi$, 0) and (0, 0)-(0,$\pi$) directions (longer Fe-Fe and shorter Fe-Fe directions) of the untwined crystal, respectively. (f) (g) Comparison of the second-derivative plot of the near-E$_F$ ARPES intensity around the $\Gamma$ point between T = 30 and 120 K. (h) Experimental band dispersion around the $\Gamma$ point at T = 30 K (blue circles) and 120 K (red circles), extracted by tracing the peak maxima of the EDCs divided by the Fermi-Dirac function. Reprinted from \cite{KNakayama}.}
\end{figure}

\begin{figure}[htbp]
\begin{center}
\includegraphics[width=1.0\columnwidth,angle=0]{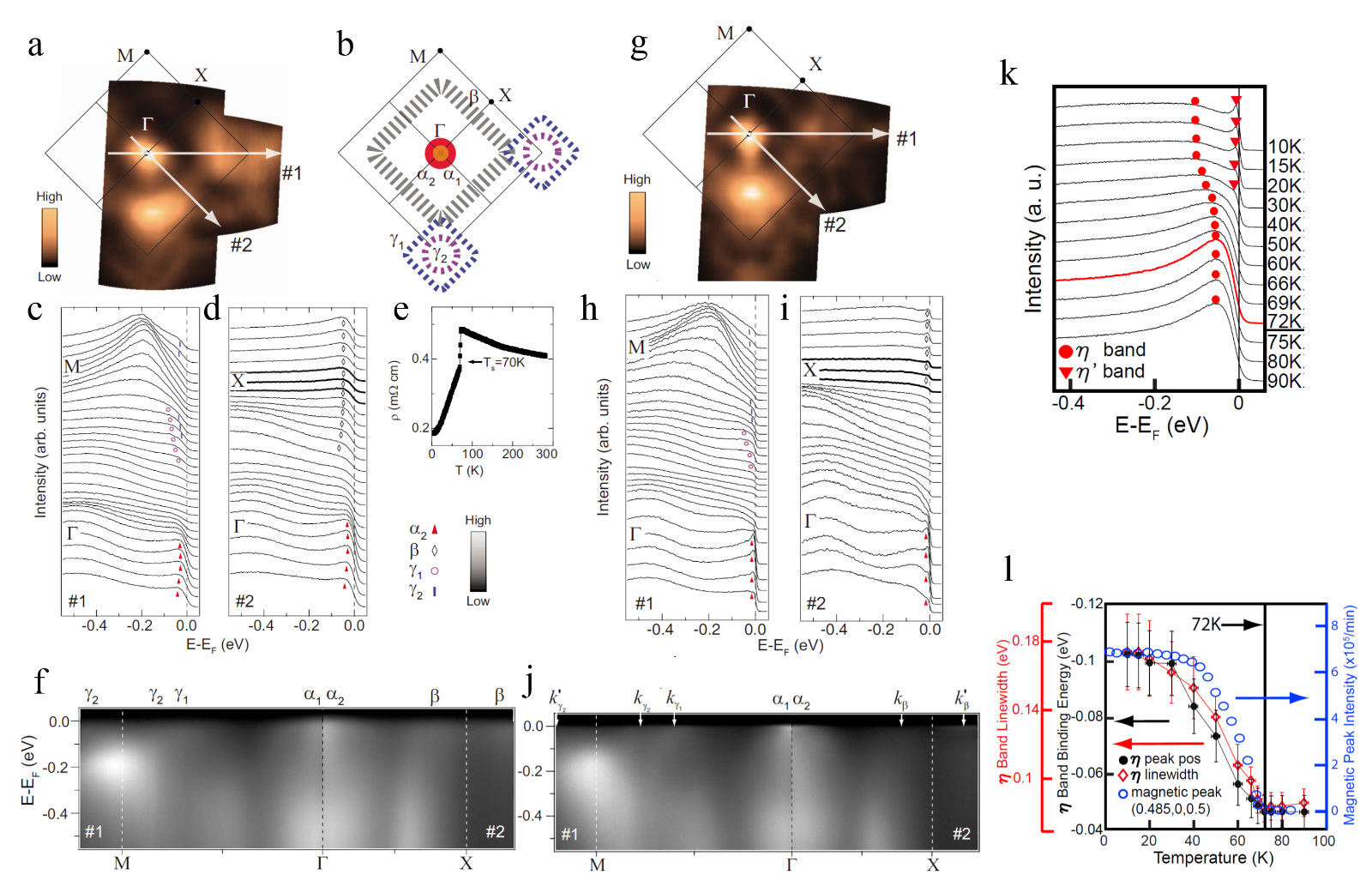}
\end{center}
\caption{\textbf{Fermi surface and band structure of FeTe.} (a)-(f) Electron structure of Fe$_{1.06}$Te in paramagnetic state above the magnetic transition. (a) Fermi surface of Fe$_{1.06}$Te measured at 135 K. (b) The spectral weight distribution around E$_F$. (c) and (d) The EDCs along cut 1 and 2 labelled in (a), respectively. (e) The temperature dependence of the resistivity for Fe$_{1.06}$Te. (f) The photoemission intensities along the M-$\Gamma$-X high-symmetry line in the paramagnetic state. (g)-(j) Electron structure of Fe$_{1.06}$Te in the magnetic state. (g) Fermi surface of Fe$_{1.06}$Te measured at 15 K. (h) and (i) The EDCs along cut 1 and cut 2 labelled in (g), respectively. (j) The photoemission intensities along the M-$\Gamma$-X high-symmetry lines in the magnetic state. Reprinted from \cite{YZhangFeTe}. (k) Plot of the EDCs at the M point at various temperatures. (l) Plot of the M band binding energy and linewidth at the M point together with the (0.5, 0, 0.5) antiferromagnetic Bragg peak intensity versus temperature. Reprinted from \cite{ZKLiuFeTe}. }
\end{figure}

\begin{figure}[htbp]
\begin{center}
\includegraphics[width=1.0\columnwidth,angle=0]{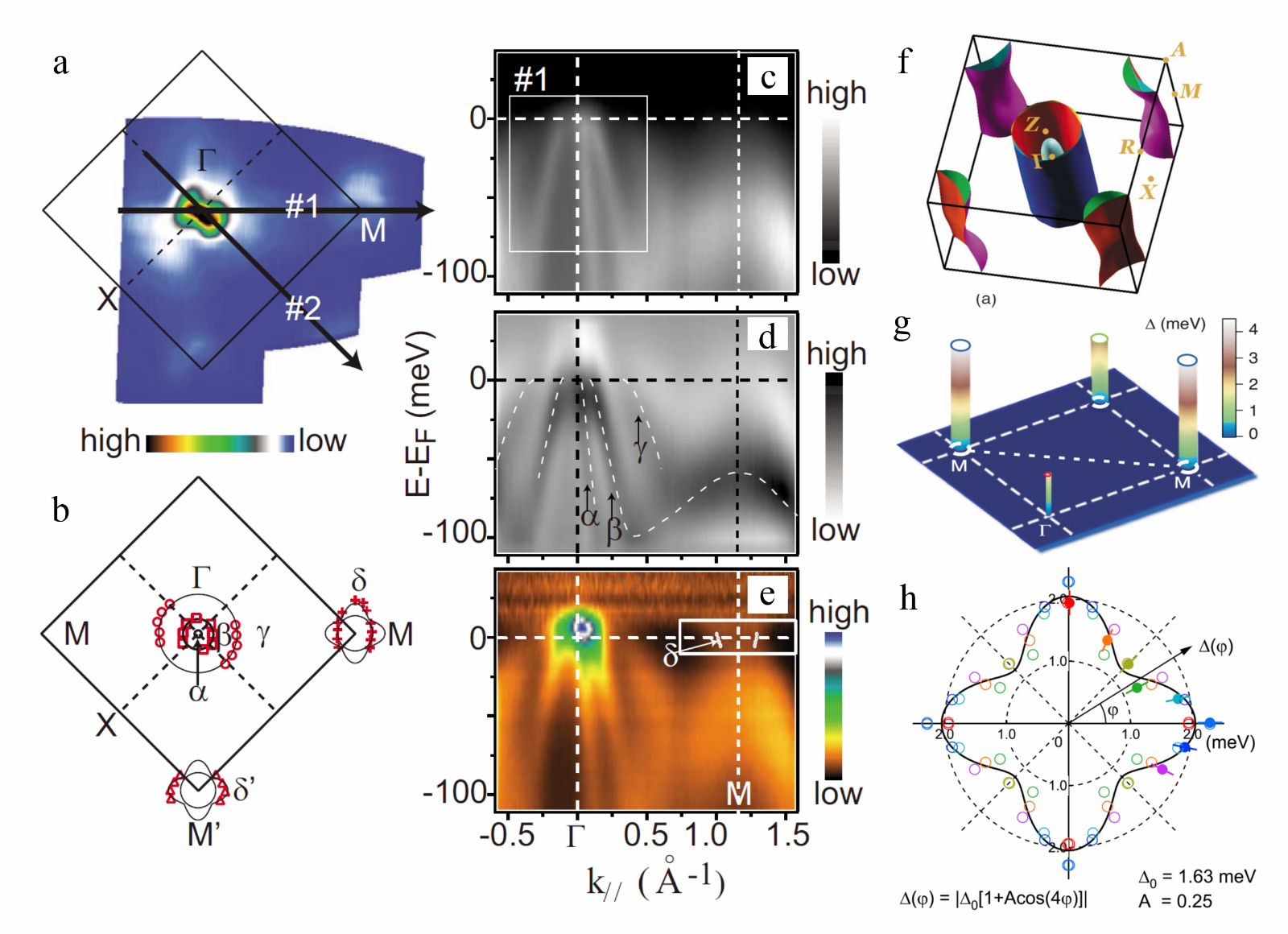}
\end{center}
\caption{\textbf{Fermi surface, band structure and superconducting gap of Fe(Te,Se).} (a) Fermi surface of Fe$_{1.04}$Te$_{0.66}$Se$_{0.34}$. (b) The Fermi surfaces are constructed based on the measured Fermi crossings. (c) The photoemission intensity along the cut 1 in the $\Gamma$-M direction and (d) its second derivative with respect to energy. (e) The data in panel c is re-plotted after dividing the angle integrated energy distribution curve. (f) The calculated Fermi surface of Fe$_{1.04}$Te$_{0.66}$Se$_{0.34}$. Reprinted from \cite{FChen}. (g) Three-dimensional representation of the superconducting gap with the Fermi surface topology of FeTe$_{0.55}$Se$_{0.45}$. Reprinted from \cite{HMiaoFeTeSe}. (h) Fermi surface-angle dependence of superconducting gap size of FeTe$_{0.6}$Se$_{0.4}$. Reprinted from \cite{KOkazakiFeTeSe} }
\end{figure}

\begin{figure}[htbp]
\begin{center}
\includegraphics[width=1.0\columnwidth,angle=0]{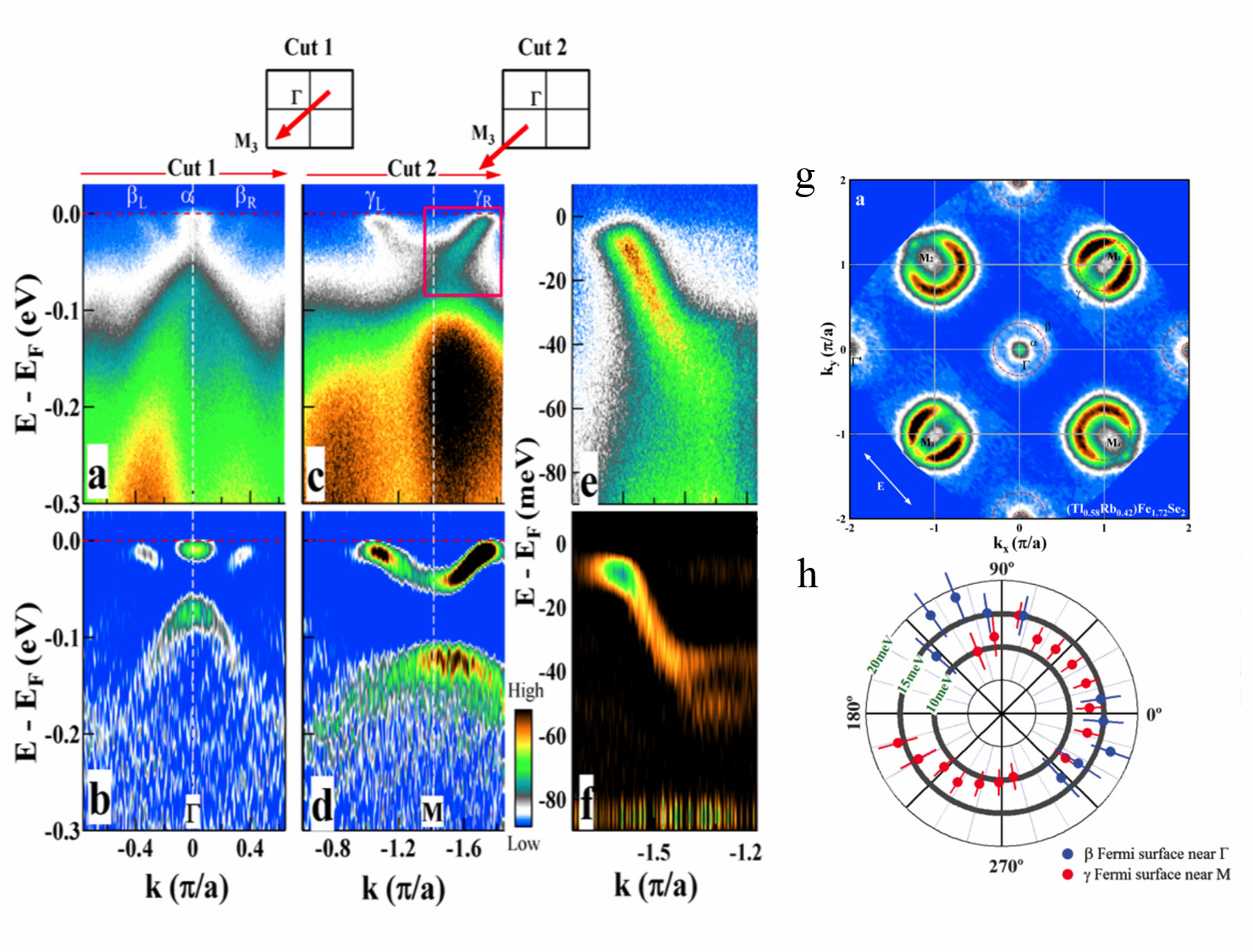}
\end{center}
\caption{\textbf{Band structure, Fermi surface and superconducting gap of  A$_x$Fe$_{2-y}$Se$_2$.} Band structure and photoemission spectra of A$_x$Fe$_{2-y}$Se$_2$ superconductors measured along two high symmetry cuts. Cut locations are illustrated in the top-left and top-right inserts. (a)(c) Measured band structure of (Tl$_{0.58}$Rb$_{0.42}$)Fe$_{1.72}$Se$_2$ along cut1 and cut2, respectively. (b)(d) Their corresponding EDC second derivative images. (e)(f) Fine measurement of the band structure in red square of (c) and its corresponding EDC second derivative images. (g) Fermi Surface of (Tl$_{0.58}$Rb$_{0.42}$)Fe$_{1.72}$Se$_2$. (h) Momentum dependent superconducting gap size for the $\beta$ Fermi surface sheet around $\Gamma$ and $\gamma$ Fermi surface sheet around M, respectively. Reprinted from \cite{DXMouPRL,DXMouReview}.
}
\end{figure}

\begin{figure}[htbp]
\begin{center}
\includegraphics[width=1.0\columnwidth,angle=0]{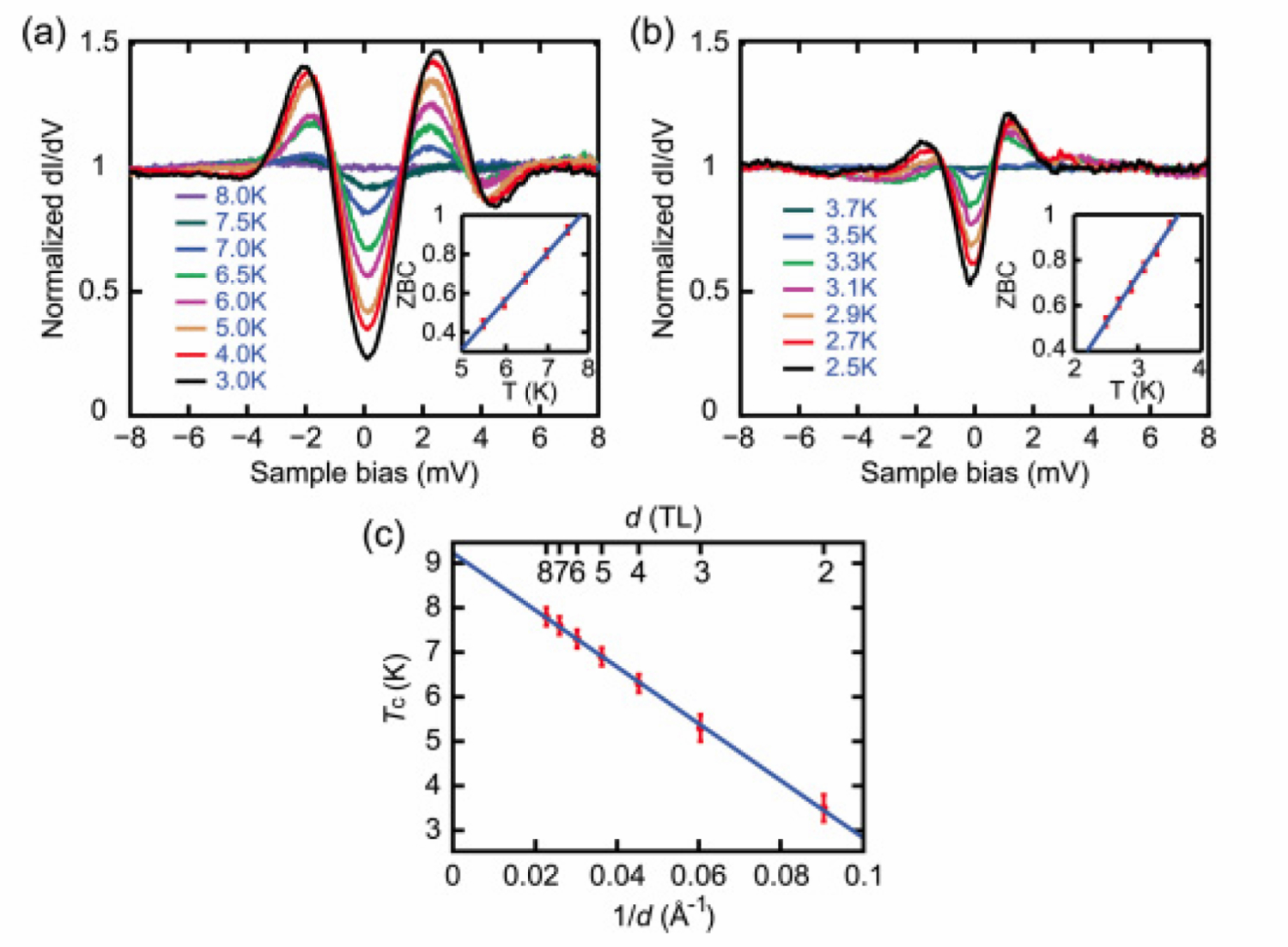}
\end{center}
\caption{\textbf{Normalized tunneling conductance spectra and T$_c$ as a function of the  film thickness in the FeSe/graphene films.} Normalized tunneling conductance spectra on (a) 8-layer and (b) 2-layer FeSe/graphene films measured at different temperatures.  (c) Superconducting transition temperature T$_c$ vs the inverse of the film thickness d.  Reprinted from \cite{CLSong}.
}
\end{figure}

\begin{figure}[htbp]
\begin{center}
\includegraphics[width=1.0\columnwidth,angle=0]{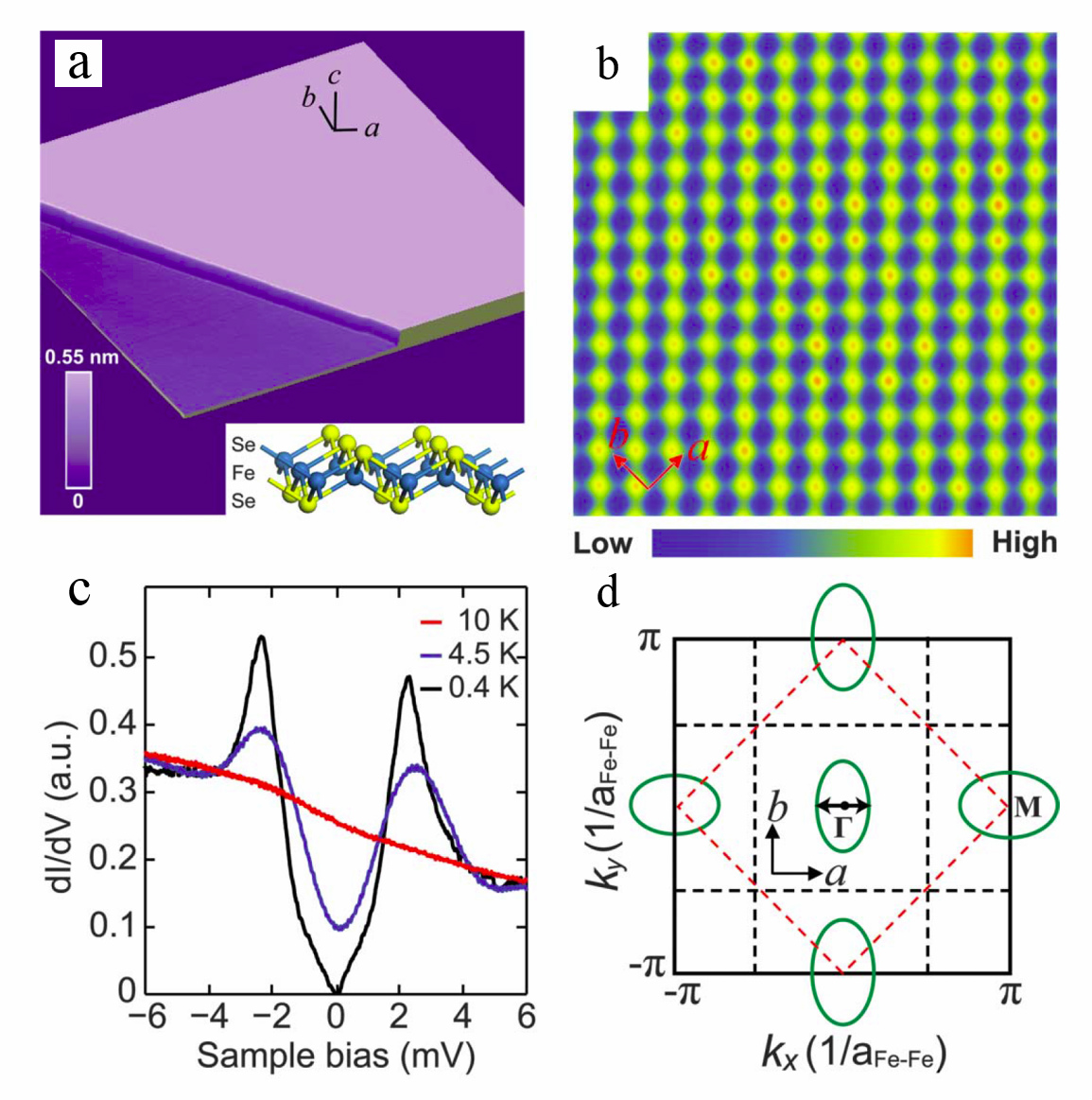}
\end{center}
\caption{\textbf{STM/STS measurements on an as-grown FeSe/graphene films about 30-layer thick.} (a) Topographic image of the FeSe/graphene film. The step height is 5.5 $\r{A}$.  (b) Atomic-resolution STM topography of the FeSe/graphene film. The bright spots correspond to the Se atoms in the top layer. a and b correspond to either of the Fe-Fe bond directions.  (c) Temperature dependence of differential conductance spectra at different temperatures. (d) Schematic of the unfolded Brillouin zone and the Fermi surface (green ellipses). The nodal lines for cosk$_x$cosk$_y$ and (cosk$_x$+cosk$_y$ ) gap functions are indicated by black and red dashed lines, respectively. Reprinted from \cite{CLSongScience}.
}
\end{figure}

\begin{figure}[htbp]
\begin{center}
\includegraphics[width=1.0\columnwidth,angle=0]{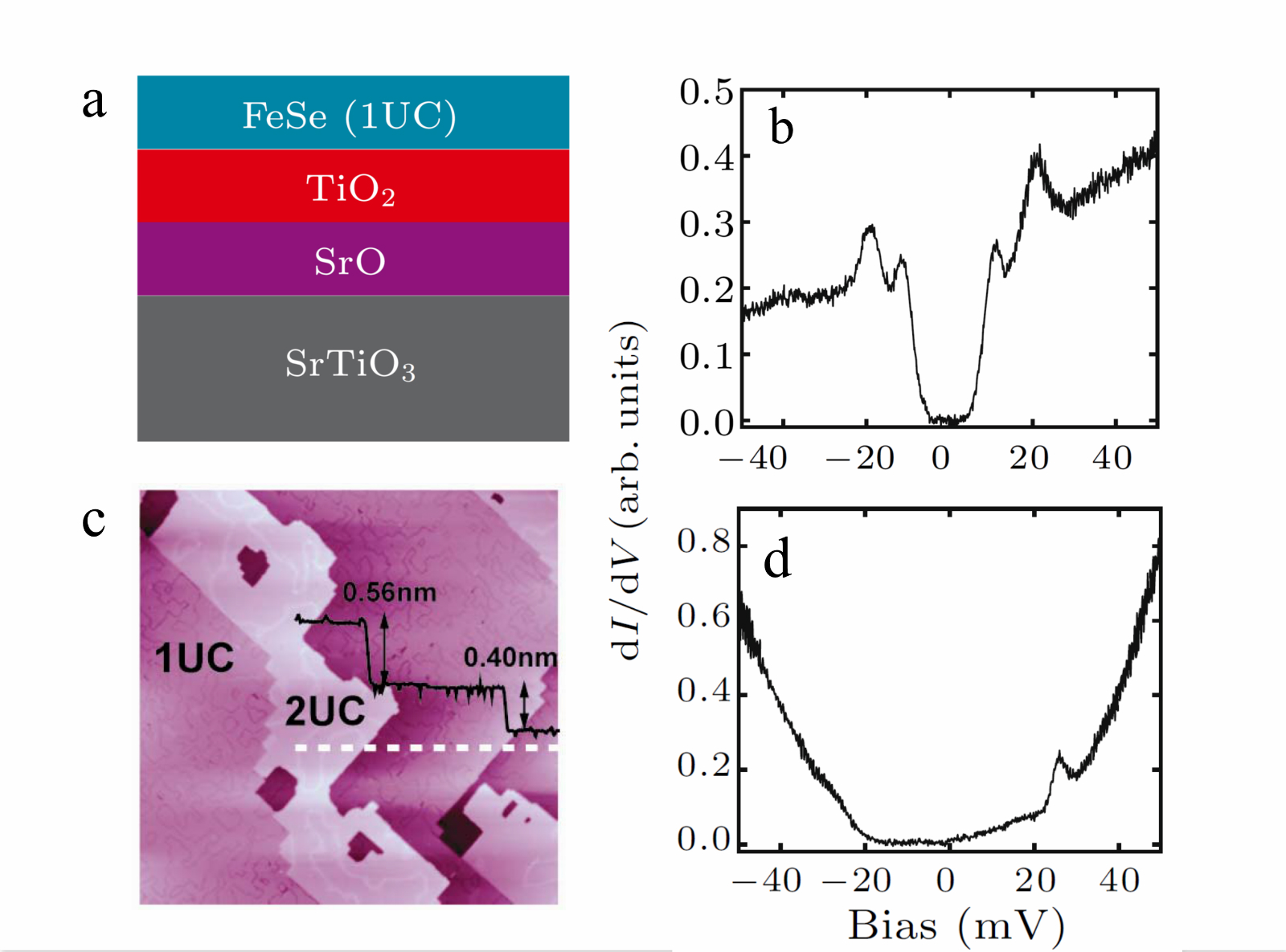}
\end{center}
\caption{\textbf{Discovery of possible high temperature superconductivity in single-layer FeSe/SrTiO$_3$ film.} (a) Schematic structure (side-view) of the FeSe films on the SrTiO$_3$ substrate along the c-axis.  (b) Tunneling spectrum taken on the single-layer  FeSe/SrTiO$_3$ film  at 4.2 K revealing the appearance of superconducting gap. Four pronounced coherence peaks appear at $\pm$20.1mV and $\pm$9mV, respectively. (c) The STM image of the FeSe film with both single-layer (1UC) and double-layer (2UC) FeSe/SrTiO$_3$ films. (d) Tunneling spectrum taken on the double-layer  FeSe/SrTiO$_3$ films. Reprinted from \cite{QYWangCPL}.}
\end{figure}

\begin{figure}[htbp]
\begin{center}
\includegraphics[width=1.0\columnwidth,angle=0]{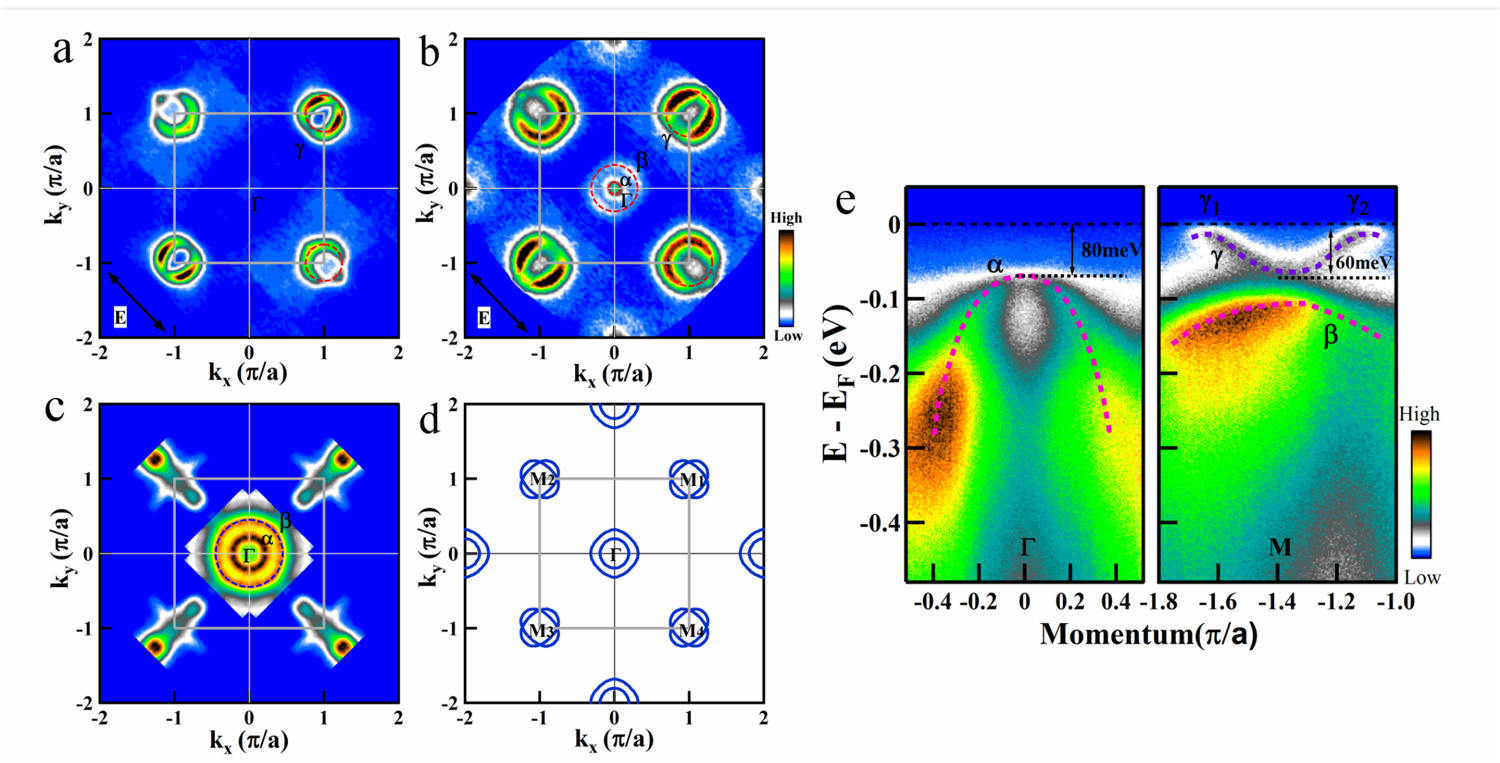}
\end{center}
\caption{\textbf{Distinct Fermi surface and band structures of the superconducting single-layer FeSe/SrTiO$_3$ Film.} (a) Fermi surface mapping of the single-layer FeSe/SrTiO$_3$ film  measured at 20 K.  (b) Fermi-surface mapping of (Tl,Rb)$_x$Fe$_{2-y}$Se$_2$ superconductor (T$_c$ = 32 K)\cite{DXMouPRL}. (c) Fermi-surface mapping of Ba$_{0.6}$K$_{0.4}$Fe$_2$As$_2$ superconductor (T$_c$ = 35 K)\cite{GDLiu}. (d) Fermi surface of $\beta$-FeSe by band-structure calculations for k$_z$ = 0 (blue thick lines)\cite{ATamai}. For convenience, the four equivalent M points are labelled as M1($\pi$, $\pi$), M2(-$\pi$, $\pi$), M3(-$\pi$, -$\pi$) and M4($\pi$, -$\pi$). (e) Band structure along the cut crossing the $\Gamma$ point (left panel) and along the cut crossing the M3 point (right panel). The pink dashed line in the left panel shows schematically a hole-like band near the $\Gamma$ point with its top at 80 meV below the Fermi level. The purple dashed line in the right panel shows schematically an electron-like band with its bottom at 60 meV below the Fermi level. Reprinted from \cite{DFLiu}.}
\end{figure}

\begin{figure}[htbp]
\begin{center}
\includegraphics[width=1.0\columnwidth,angle=0]{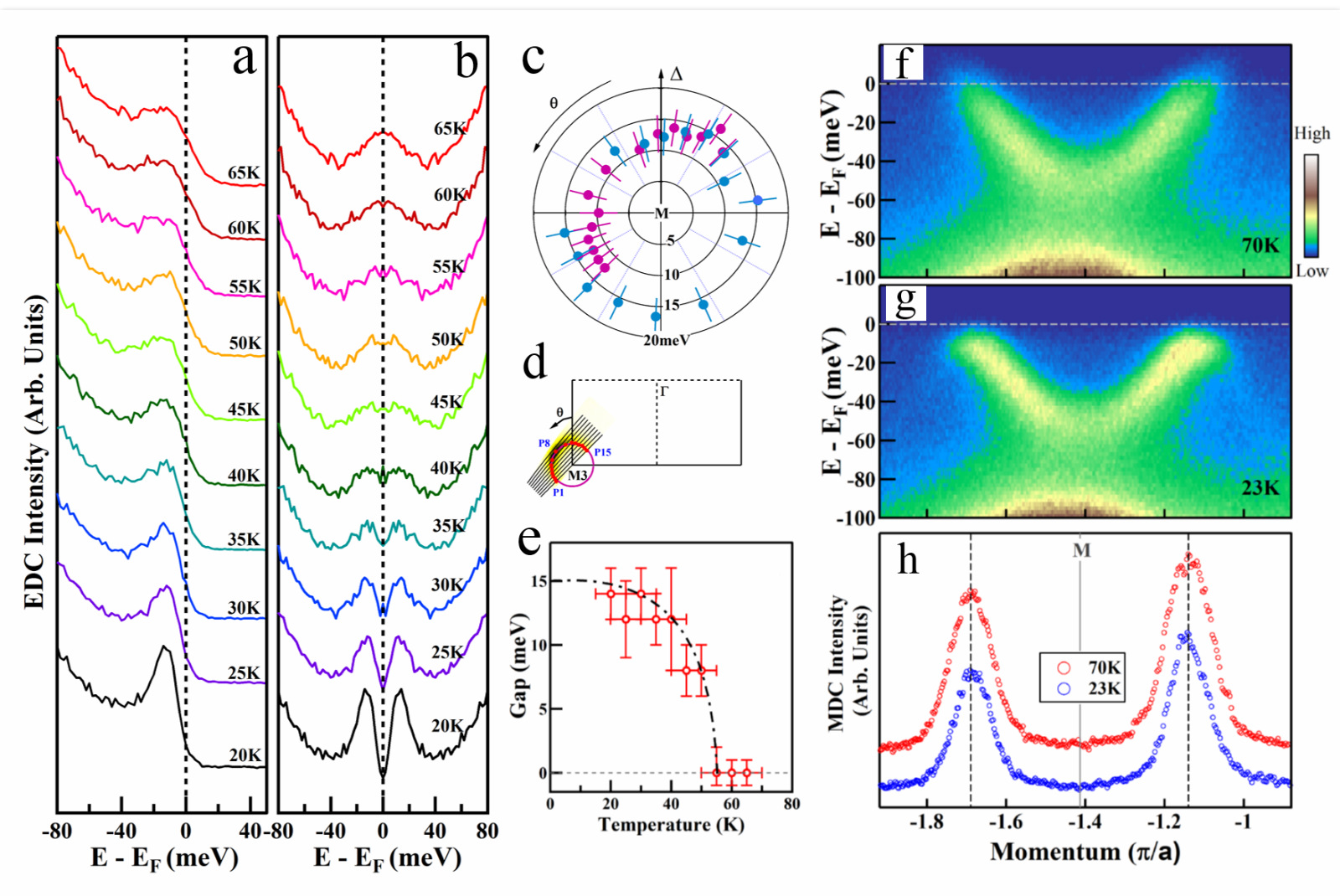}
\end{center}
\caption{\textbf{Temperature and momentum dependence of the superconducting gap in the superconducting single-layer FeSe/SrTiO$_3$ film.} (a) Photoemission spectra (EDCs) at the Fermi crossings of the electron-like $\gamma$ Fermi surface near M  and their corresponding symmetrized spectra (b) measured at different temperatures. (c) Momentum dependence of the superconducting gap along the $\gamma$ Fermi surface. (d) The Fermi surface mapping near M3 and the corresponding Fermi crossings. The violet circle represents the Fermi surface. The red solid circles represent measured Fermi momenta that are labeled as P1 to P15. (e) Temperature dependence of the superconducting gap. Reprinted from \cite{DFLiu}. (f) (g) Band structure along $\Gamma$-M cut measured at 70 and 23 K, respectively. The corresponding MDCs (momentum distribution curves) at the Fermi level for the two measurements are shown in (h).  The two MDC peaks show little change in their positions above and below the gap opening temperature. Reprinted from \cite{SLHe}.}
\end{figure}

\begin{figure}[htbp]F
\begin{center}
\includegraphics[width=1.0\columnwidth,angle=0]{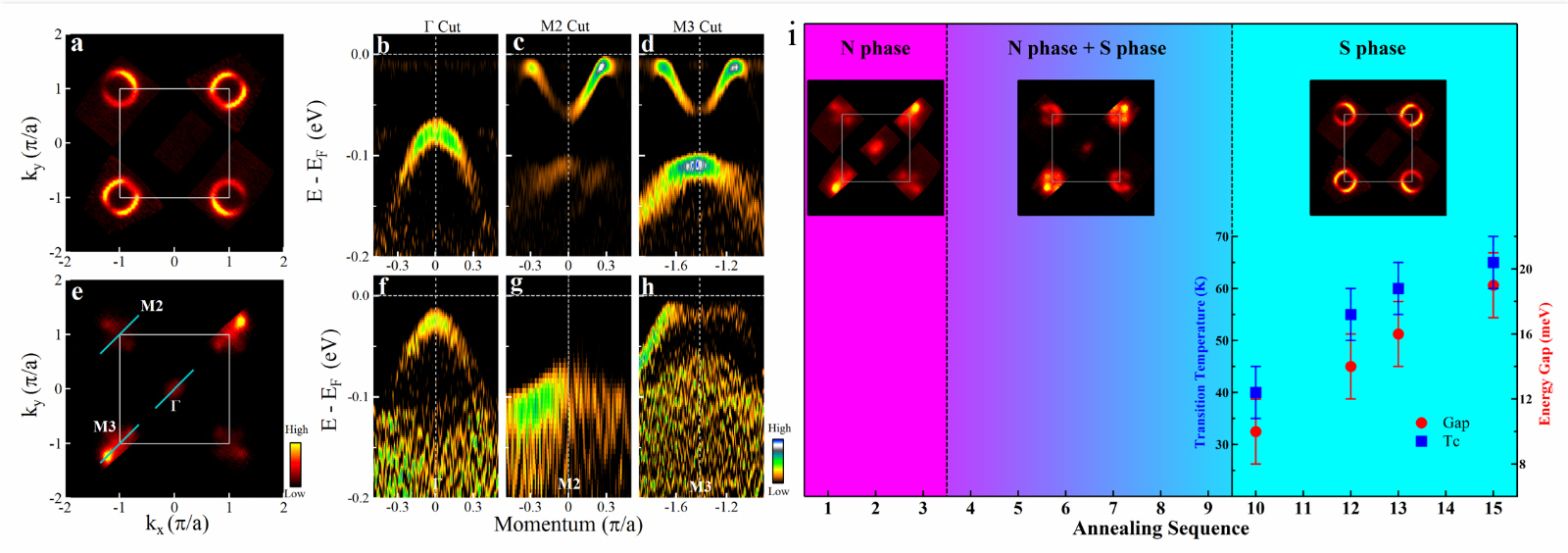}
\end{center}
\caption{\textbf{Electronic structure of the S phase and the N phase and phase diagram of the single-layer FeSe/SrTiO$_3$ film with vacuum annealing.} (a) Fermi surface of the S phase of the single-layer FeSe/SrTiO$_3$ film. (b-d) Corresponding band structure of the S phase along the $\Gamma$, M2 and M3 cuts, respectively. (e) Fermi surface of the N phase of the single-layer FeSe/SrTiO$_3$ film. (f-h), Corresponding band structure of the annealed single-layer FeSe/SrTiO$_3$ film along the $\Gamma$, M2 and M3 cuts, respectively. The band structures shown in (b,c,d) and (f,g,h) are second derivative of the original band with respect to energy. (i) Schematic phase diagram of the single-layer FeSe film during the annealing process. The inset shows the Fermi surface corresponding to each stage. Reprinted from \cite{SLHe}.}
\end{figure}

\begin{figure}[htbp]
\begin{center}
\includegraphics[width=1.0\columnwidth,angle=0]{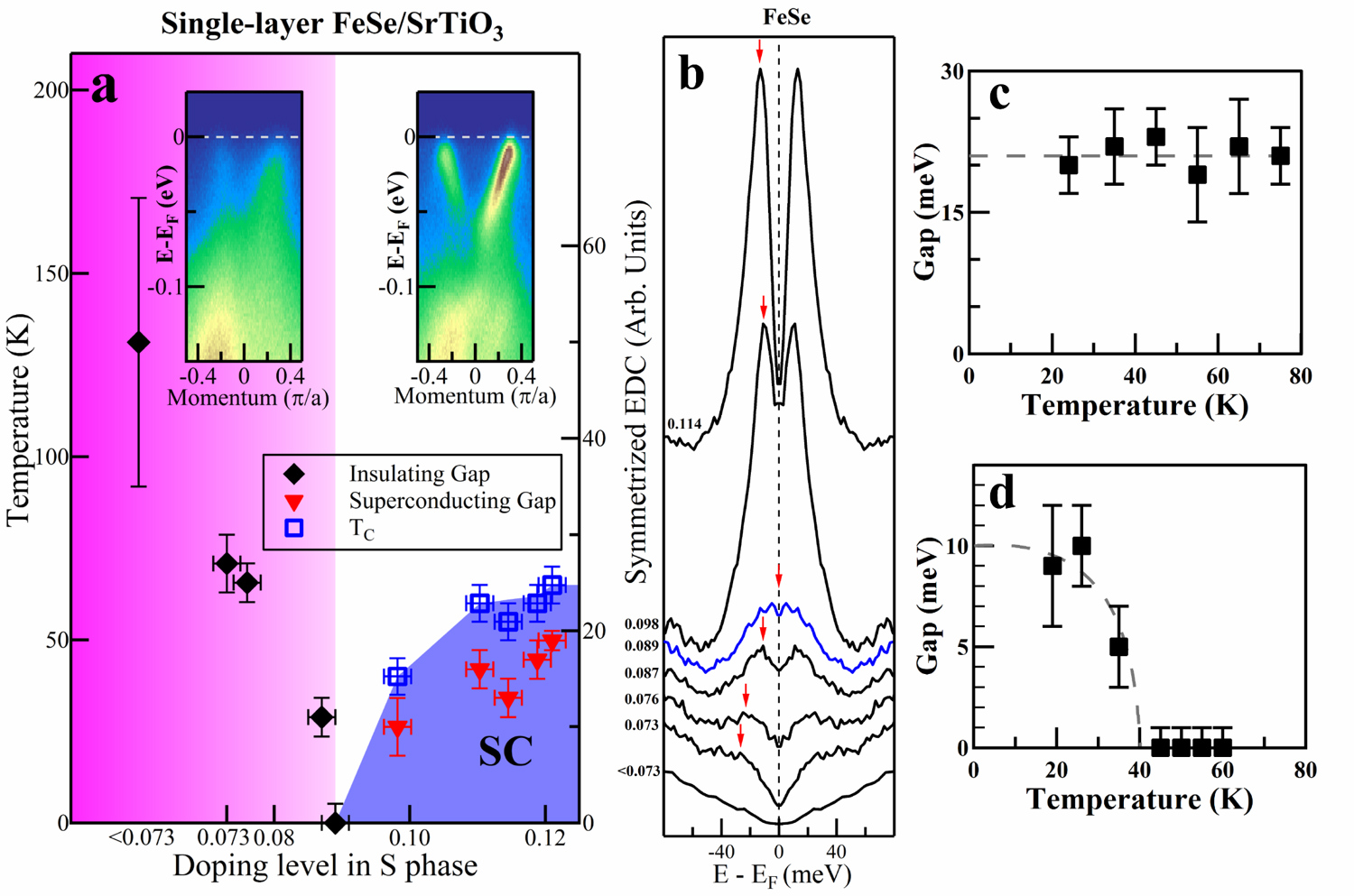}
\end{center}
\caption{\textbf{Evidence of an insulator-superconductor transition in the single-layer FeSe/SrTiO$_3$ film.} (a) Phase diagram of the S phase in the single-layer FeSe/SrTiO$_3$ film that shows the decrease of the insulating energy gap (black solid diamond) with increasing doping at low doping side and the increase of the superconducting gap (red solid triangle) and the corresponding superconducting transtion temperature T$_c$ (blue empty square) with increasing doping at high doping side. There is an insulator-superconductor transition near $\sim$0.09 doping level. (b) Doping-evolution of the symmetrized EDCs at a Fermi momentum k$_F$ along the cut crossing M2 point. (c) (d) The variation of the energy gap at different temperatures at low doping level and high doping level, respectively. Reprinted from \cite{JFHe}.}
\end{figure}

\begin{figure}[htbp]
\begin{center}
\includegraphics[width=1.0\columnwidth,angle=0]{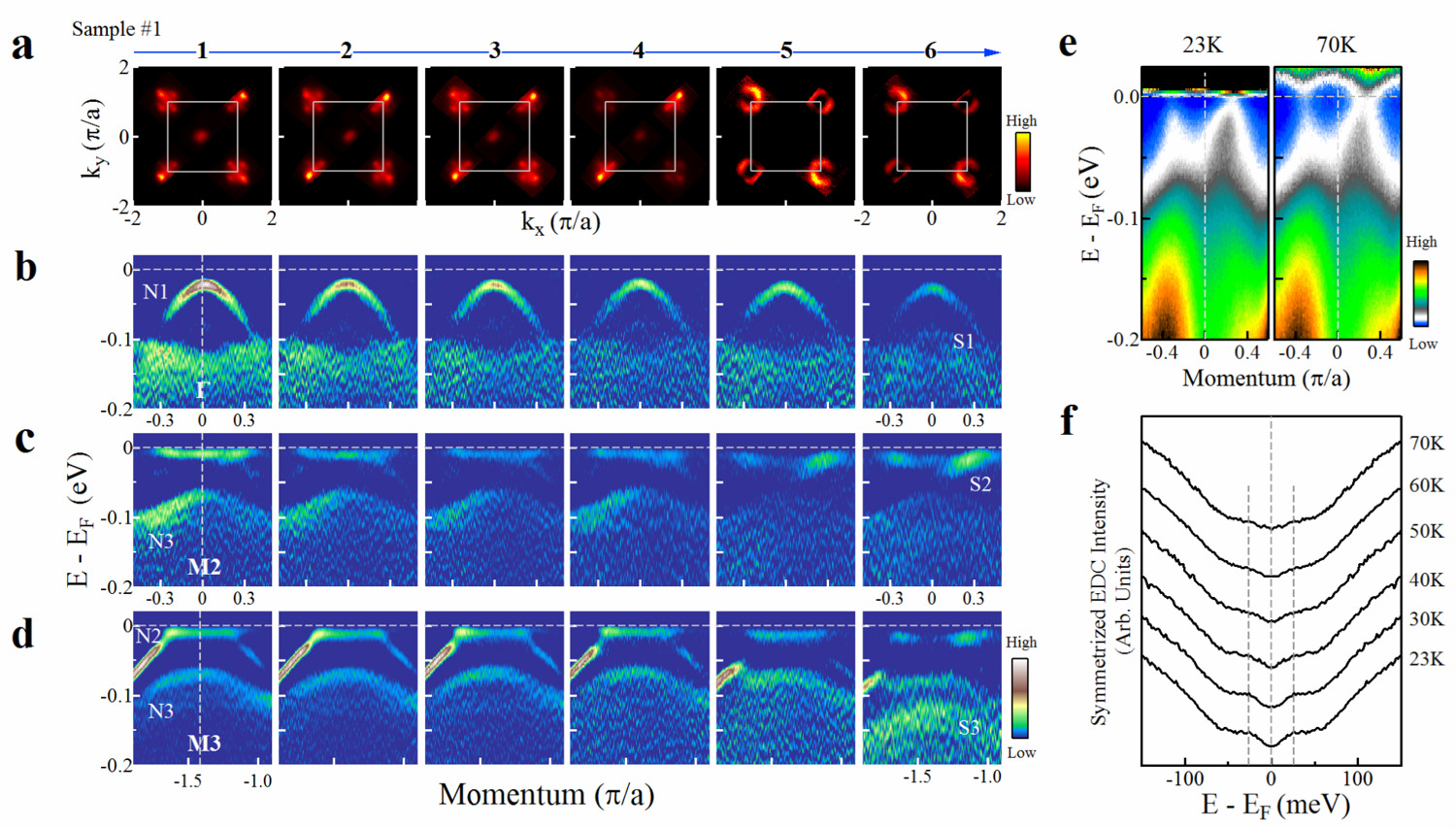}
\end{center}
\caption{\textbf{Fermi surface and band structure evolution of the double-layer FeSe/SrTiO$_3$ film annealed at a constant annealing temperature of 350$^{\circ}$C in vacuum for different times.} (a) Fermi surface and band structure evolution of the double-layer FeSe/SrTiO$_3$ film annealed at a constant annealing temperature of 350 $^{\circ}$C in vacuum for different times. The sequences 1 to 6 correspond to an accumulative time of 15, 30.5, 46.5, 66.5, 87.5, and 92.5 hours, respectively. (a) Fermi surface evolution as a function of the annealing time. (b-d) Band structure evolution with annealing time for the momentum cuts across $\Gamma$ (b), M2 (c) and M3 (d). (e) Band structure of the annealed double-layer FeSe/SrTiO$_3$ film (corresponding to the sequence 6) measured at 23 K (left panel) and 70 K (right panel). The photoemission images are divided by the corresponding Fermi distribution function to highlight opening or closing of an energy gap. (f) Corresponding symmetrized EDCs on the Fermi momentum measured at different temperatures. Reprinted from \cite{XLiu}.}
\end{figure}

\begin{figure}[htbp]
\begin{center}
\includegraphics[width=1.0\columnwidth,angle=0]{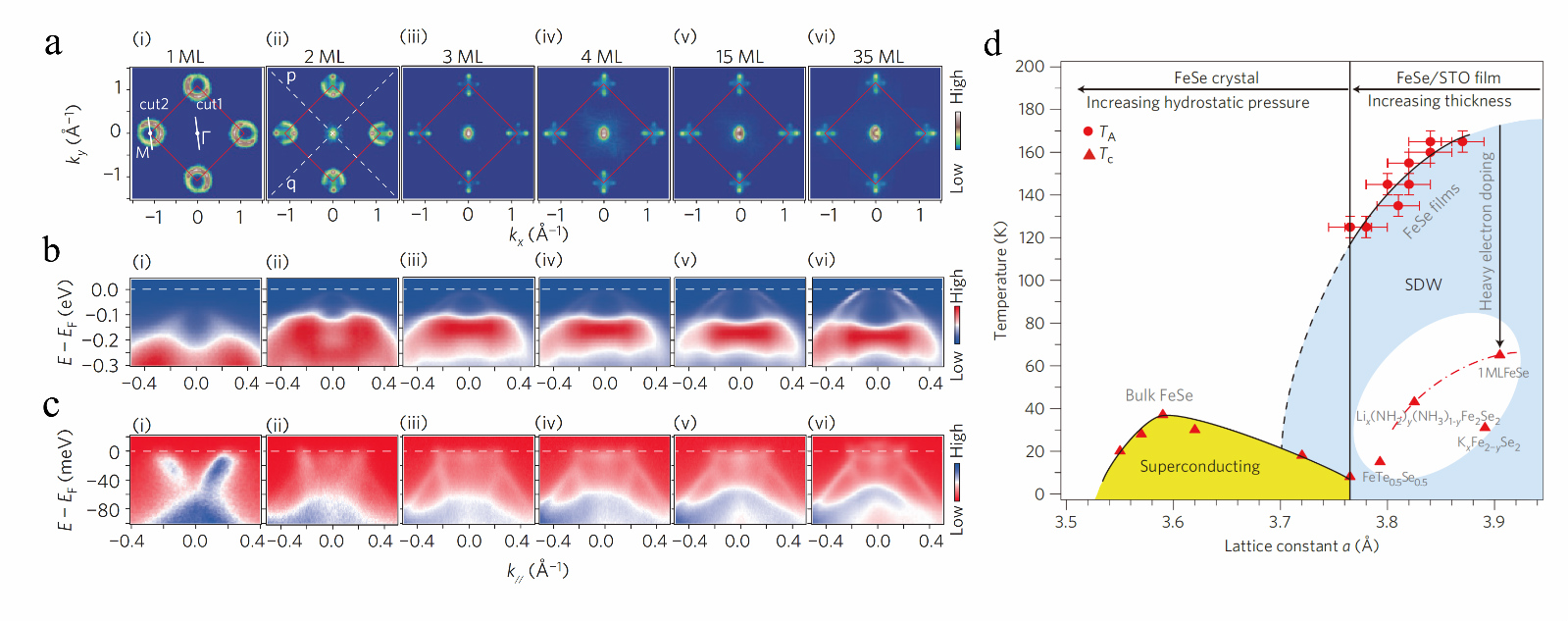}
\end{center}
\caption{\textbf{Layer dependent Fermi surface and band structure in the FeSe/SrTiO$_3$ films.} (a) The thickness dependence of the Fermi surface as represented by the photoemission intensity map at the Fermi energy at 30 K. (b) (c) The thickness dependence of band structure along the cut crossing $\Gamma$ point and M point respectively. (d) Phase diagram of the FeSe-related systems. Values of T$_c$ and T$_A$ for FeSe are plotted against the lattice constant. The dashed line represents the extrapolated values of T$_A$, suggesting the existence of spin-density-wave (SDW) order in bulk FeSe under pressure. Values of T$_c$ for other iron selenides are also plotted in the elliptical region. Reprinted from \cite{STan}.}
\end{figure}

\begin{figure}[htbp]
\begin{center}
\includegraphics[width=1.0\columnwidth,angle=0]{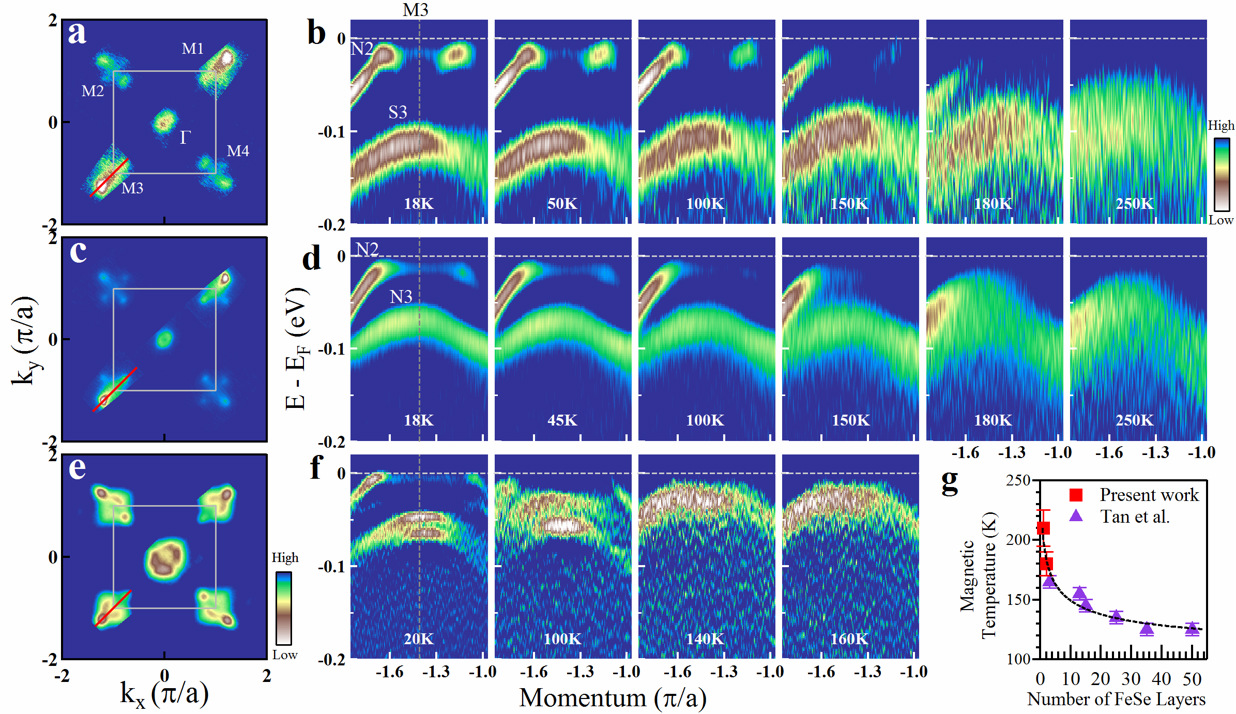}
\end{center}
\caption{\textbf{Temperature dependence of the band structure of the N phase in the single-layer and double-layer FeSe/SrTiO$_3$ films.} (a),(c) and (e) are the Fermi surface of single-layer FeSe/SrTiO$_3$ film, double-layer FeSe/SrTiO$_3$ film and BaFe$_2$As$_2$, respectively. (b), (d) and (f) Temperature dependence of the band structure measured along the M3 cut for the single-layer FeSe/SrTiO$_3$ film, double-layer FeSe/SrTiO$_3$ film and BaFe$_2$As$_2$, respectively.  (g) The transition temperature as a function of the number of FeSe layers in the FeSe/SrTiO$_3$ films. Here the transition temperature for the single-layer and double-layer FeSe films (red squares) is determined from the temperature dependence of the band structure where the hole-like bands begin to disappear. The temperature for other films (violet triangles) with more FeSe layers is taken from \cite{STan}. Reprinted from \cite{XLiu}.}
\end{figure}

\begin{figure}[htbp]
\begin{center}
\includegraphics[width=1.0\columnwidth,angle=0]{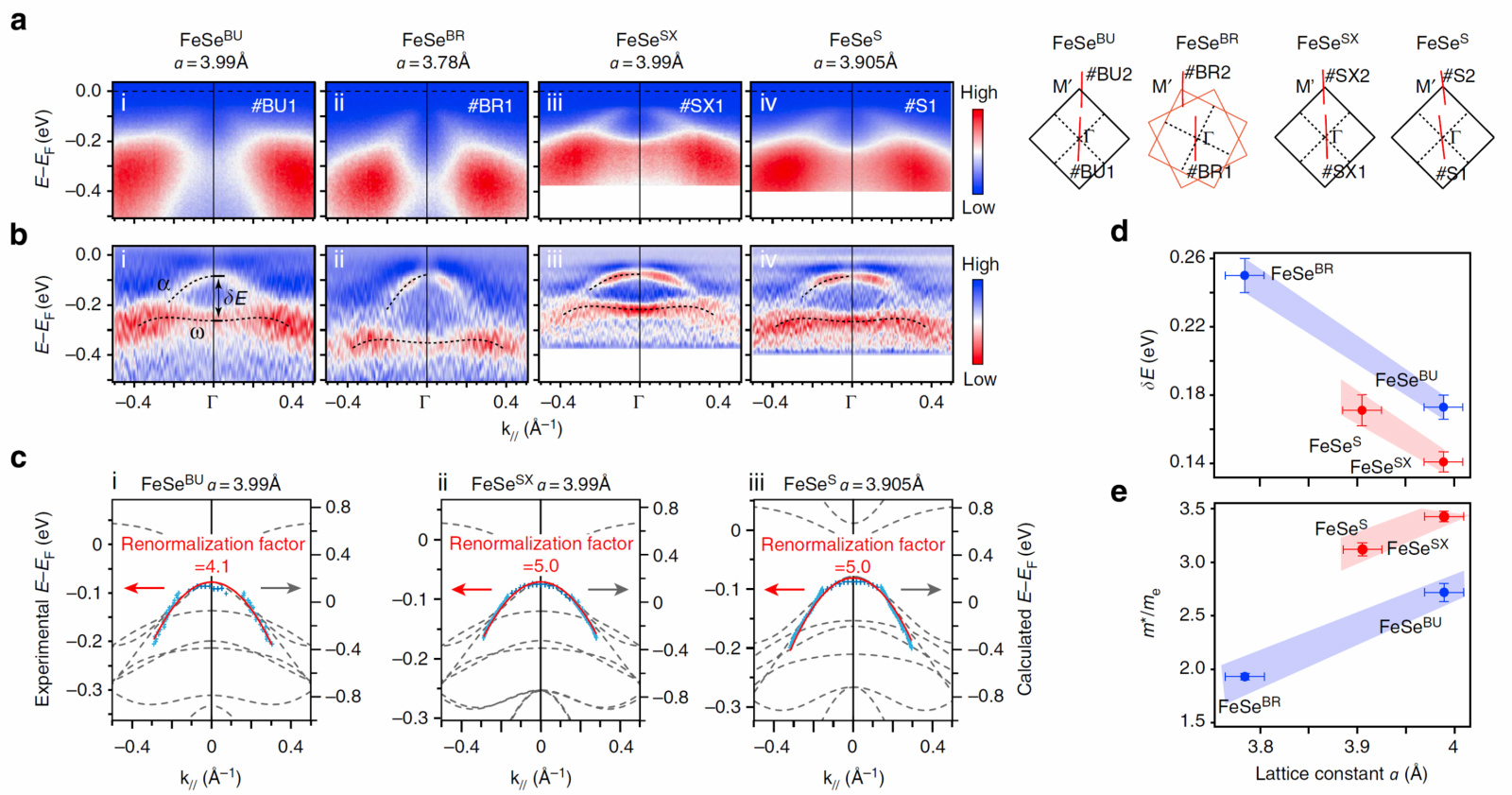}
\end{center}
\caption{\textbf{Band structure of the single-layer FeSe films grown on various substrates.} (a,b) The photoemission intensity along $\#$BU1, $\#$BR1, $\#$SX1 and $\#$S1 across $\Gamma$, and the corresponding second derivative with respect to energy to highlight the dispersions  for FeSe$^{BU}$, FeSe$^{BR}$, FeSe$^{SX}$ and FeSe$^S$, respectively. For the definition of different substrates, refer to \cite{RPengNC}.  (c) Comparison of the dispersion of band $\alpha$ from ARPES data and the DFT calculated band structures along the $\Gamma$-M' direction. (d) The energy separations between the $\alpha$ and $\omega$ bands at $\Gamma$ versus a. (e) Effective mass m$^*$ of band $\alpha$ as a function of in-plane lattice constant a, where m$_e$ is the free-electron mass. Reprinted from \cite{RPengNC}.}
\end{figure}

\begin{figure}[htbp]
\begin{center}
\includegraphics[width=1.0\columnwidth,angle=0]{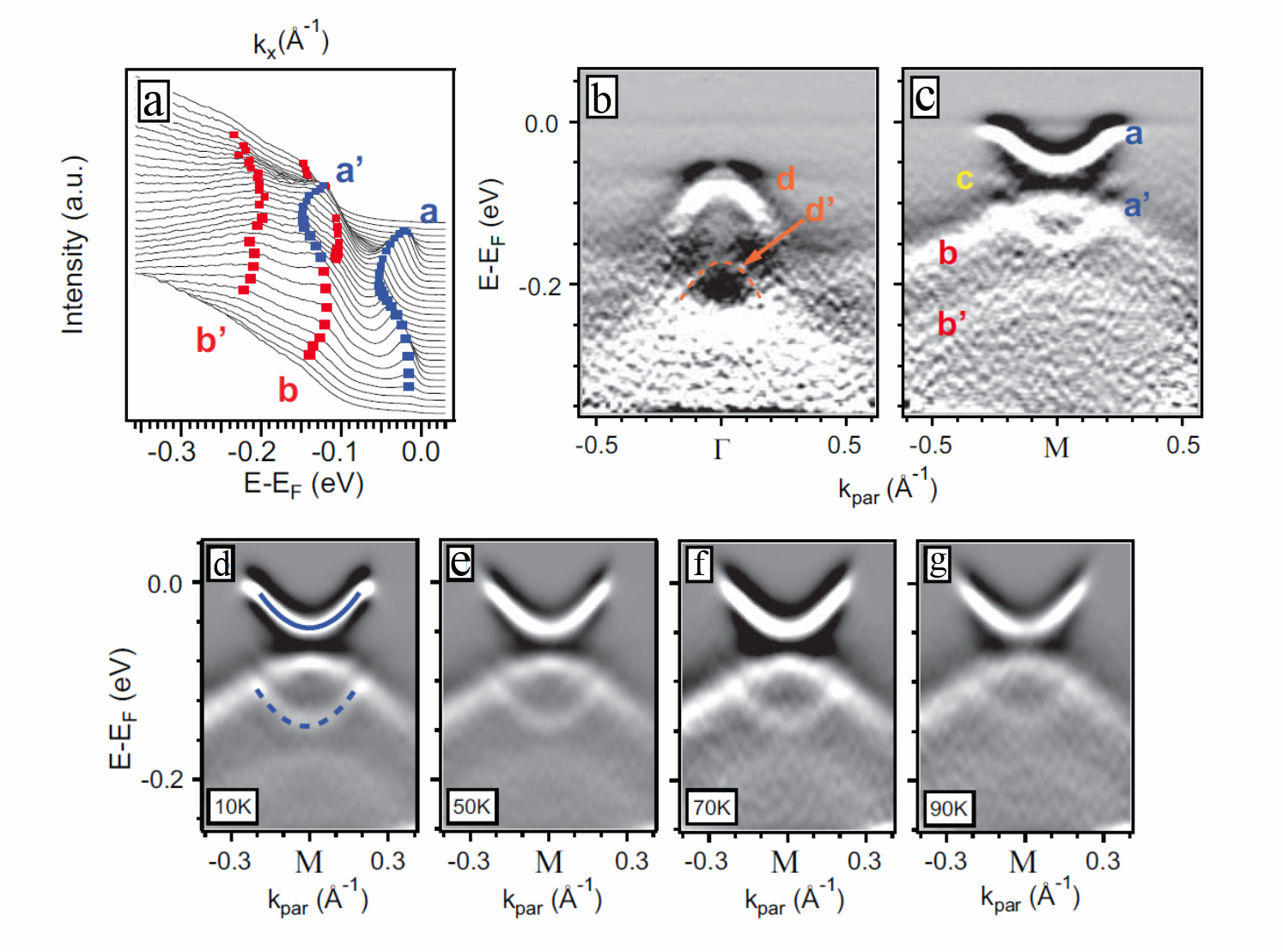}
\end{center}
\caption{\textbf{Replica bands observed in the single-layer FeSe/SrTiO$_3$ films and their temperature dependence.} (a) EDCs at M shown as a waterfall plot, with markers indicating band peaks. (b) (c) Second derivatives in energy of the high symmetry cuts from $\Gamma$ cut and M cut. An additional weaker replica, labelled c, can now be seen at M in (c), sitting 50 meV below a, and at the $\Gamma$ point in (b) we see the hole band and a corresponding replica, labelled d and d¡ä, respectively. (d-g) Temperature dependence of the replica bands, which persist at temperatures higher than the gap-opening temperature. Reprinted from \cite{JJLee}.}
\end{figure}

\begin{figure}[htbp]
\begin{center}
\includegraphics[width=1.0\columnwidth,angle=0]{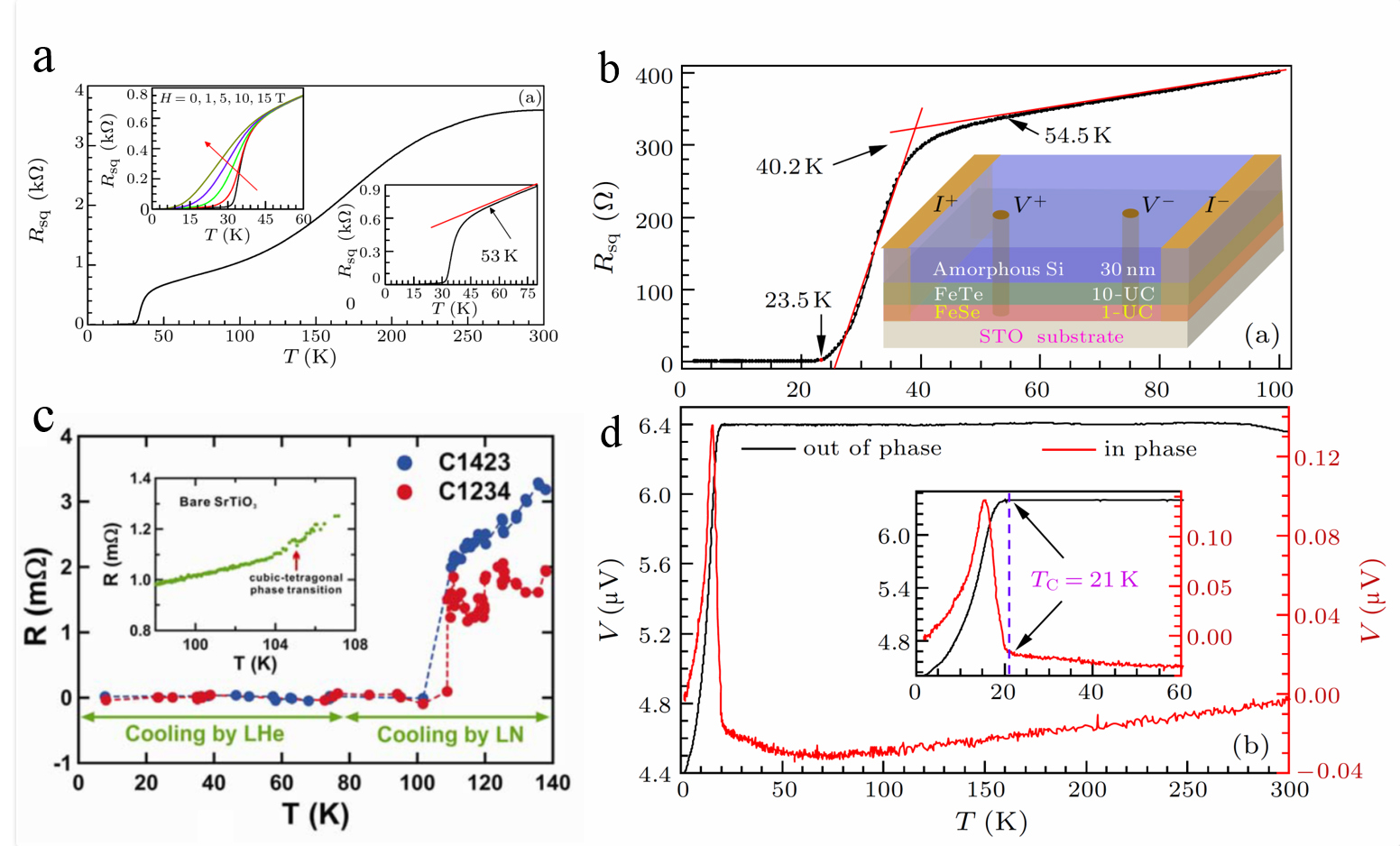}
\end{center}
\caption{\textbf{Transport and magnetic measurements of the superconductivity in the FeSe/SrTiO$_3$ films.} (a) Temperature dependence of square resistivity (R$_{sq}$) of a 5-layer FeSe/SrTiO$_3$ film from 0 to 300 K. Upper inset: R$_{sq}$$¨C$T curves at various magnetic fields along the c-axis. Lower inset: the R$_{sq}$$¨C$T curve from 0 to 80 K\cite{QYWangCPL}. (b)The temperature dependence of resistance under zero field, showing T$^{onset}$$_c$ =40.2K and T$^{zero}$$_c$ =23.5 K. Inset: a schematic structure for the transport measurements in the heterostructure of 30nm amorphous Si/(10-layer)FeTe/(1-layer)FeSe/SrTiO$_3$\cite{WHZhangCPL}. (c) Temperature dependence of the resistance in situ measured on a single-layer FeSe/SrTiO$_3$ film\cite{JFGe}. The inset shows the temperature dependence of resistance taken on a bare SrTiO$_3$ surface. (d) The diamagnetic response measured by a two-coil mutual inductance system. Inset: the data near the superconducting transition in an amplified view showing the formation of diamagnetic screening at 21 K\cite{WHZhangCPL}.}
\end{figure}
\clearpage

\begin{figure}[htbp]
\begin{center}
\includegraphics[width=0.6\columnwidth,angle=0]{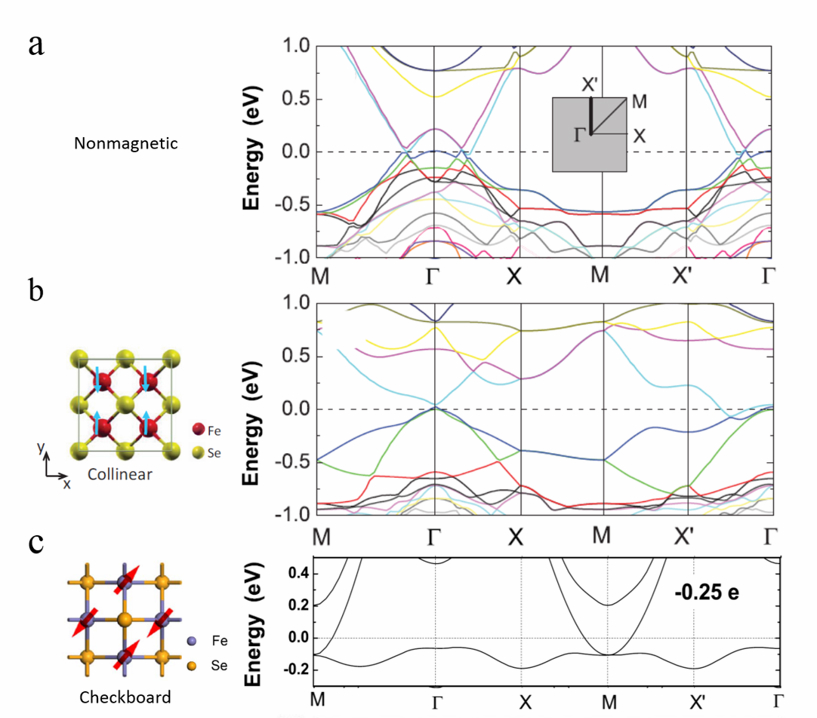}
\end{center}
\caption{\textbf{Band structure Calculations of single-layer FeSe film with different magnetic orders.} (a) and (b) The calculated band structure of single-layer FeSe/SrTiO$_3$ film in nonmagnetic state and collinear antiferromagnetic state respectively\cite{KLiu}. (c) The calculated band structure of single-layer FeSe film in checkboard antiferromagnetic state\cite{FWZheng}.}
\end{figure}

\end{document}